\documentclass[lettersize,journal]{IEEEtran}
\usepackage{amsmath,amsfonts}
\usepackage{algorithmic}
\usepackage{algorithm}
\usepackage{array}
\usepackage[caption=false,font=normalsize,labelfont=sf,textfont=sf]{subfig}
\usepackage{textcomp}
\usepackage{stfloats}
\usepackage{url}
\usepackage{verbatim}
\usepackage{graphicx}
\usepackage{cite}
\usepackage{multirow}
\usepackage{multicol}
\usepackage{lipsum}
\usepackage[usestackEOL]{stackengine}

\usepackage[dvipsnames]{xcolor}
\begin{document}
\title{Variable Resolution Pixel Quantization for Low Power Machine Vision Application on Edge}
\author{Senorita Deb, Sai Sanjeet, Prabir Kumar Biswas, and Bibhu Datta Sahoo
\thanks{Senorita Deb and Prabir Kumar Biswas are with the Department of Electronics and Electrical Communication Engineering, Indian Institute of Technology Kharagpur, India. \\
\indent Sai Sanjeet and Bibhu Datta Sahoo are with Department of Electrical Engineering, University at Buffalo, Buffalo, NY-14260. e-mail: bsahoo@ieee.org.}}
\maketitle
\begin{abstract}
This work describes an approach towards pixel quantization using variable resolution which is made feasible using image transformation in the analog domain. The main aim is to reduce the average bits-per-pixel (BPP) necessary for representing an image while maintaining the classification accuracy of a Convolutional Neural Network (CNN) that is trained for image classification. The proposed algorithm is based on the Hadamard transform that leads to a low-resolution variable quantization by the analog-to-digital converter (ADC) thus reducing the power dissipation in hardware at the sensor node. Despite the trade-offs inherent in image transformation, the proposed algorithm achieves competitive accuracy levels across various image sizes and ADC configurations, highlighting the importance of considering both accuracy and power consumption in edge computing applications. The schematic of a novel $1.5$ bit ADC that incorporates the Hadamard transform is also proposed. A hardware implementation of the analog transformation followed by software-based variable quantization is done for the CIFAR-10 test dataset. The digitized data shows that the network can still identify transformed images with a remarkable $90\%$ accuracy for $3$-BPP transformed images following the proposed method.
\end{abstract}

\begin{IEEEkeywords}
Analog-to-Digital Converters, Hadamard Transform,  Machine vision, Convolutional Neural Networks (CNN).
\end{IEEEkeywords}
\IEEEpeerreviewmaketitle
\section{Introduction}
\label{sec: introduction}
\begin{figure}[h]
\centering
\includegraphics[scale=0.25]{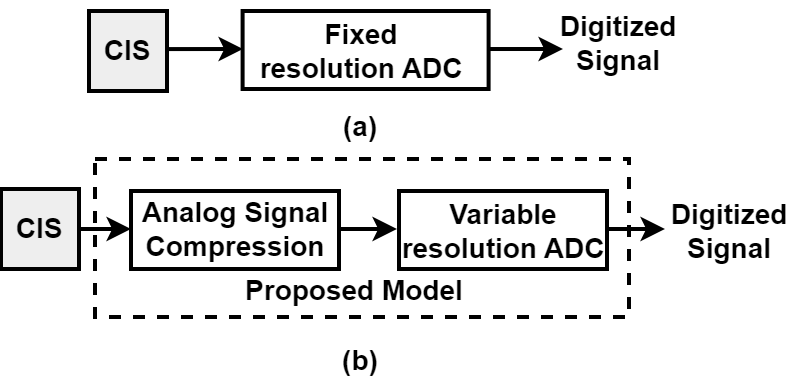}
\vspace{-1em}
\caption{\textcolor{black}{(a) Conventional ADC architecture (b) Proposed Architecture for Data Acquisition from Edge Sensor.}} \vspace{-1em}
\label{fig: proposed}
\end{figure}

In an era of computationally demanding applications with limited battery capacity \cite{mach2017mobile} developing energy-efficient mobile applications becomes increasingly challenging. In order to ensure the sustainability and longevity of Internet of Things (IoT) deployments, it is essential to minimize power consumption in edge devices \cite{lee201624} without compromising performance. With energy harvesting providing only limited available power, the design of edge computing devices that consume minimal power and occupy less chip area becomes an urgent need \cite{sadhu2013analysis}. Efficient management of power dissipation by edge sensors, optimization of data transmission bandwidth from edge devices to end computing devices, enhancement of throughput, and expansion of storage capacity in processors are crucial factors that demand attention. Addressing these aspects is essential for the development of efficient edge devices that can meet the requirements of modern computational applications. Therefore, quantization of the sensor output while consuming minimal power and reduction of I/O bandwidth without degrading its quality are the avenues that can be optimized. Image compression techniques by applying various orthogonal transforms like Haar transform, Walsh-Hadamard transform (WHT) also referred to as Hadamard Transform (HT), Slant transform, Discrete Cosine transform (DCT), etc. have been reported in literature for energy efficiency  \cite{azar2019energy, uthayakumar2020highly, krishnaraj2020deep,djelouat2018system}. Although, lossless compression techniques are preferred for lot of applications, machine vision applications can achieve very good accuracy with images compressed using lossy compression techniques like HT, Discrete Cosine Transform (DCT), etc. \cite{yang2023introduction, lungisani2022image}.\\
\indent Although, quantizing the images to lower resolution could lead to significant compression and energy efficiency, the accuracy of machine vision applications get severely compromised \cite{Sanjeet2023} with image pixels quantized to very low resolution. \textcolor{black}{Fig. \ref{fig: proposed}(a) shows the traditional image readout system where the analog signal from the CMOS Image Sensor (CIS) is passed to a quantizer with fixed resolution to obtain the digital signal. This paper, as shown in Fig. \ref{fig: proposed}(b), proposes a one dimensional-Hadamard transform (1D-HT), implemented in analog domain, of analog signal from the CIS before digitization by the ADC.} This leads to a low-resolution variable quantization by the ADC thus reducing the power dissipation in hardware at the sensor node. This digitized data is transmitted from the sensor to the processor, where an inverse HT is done to reconstruct the image. The reconstructed image is passed to a CNN for image classification.
The proposed method of pixel quantization, {\em i.e.,} analog 1D-HT followed by variable resolution quantization of the image pixel, when compared with regular pixel quantization with lower bit-precision\footnote{Image pixel quantization by simply lowering the ADC resolution will be referred to as {\em baseline-quantization}.}, performs better in terms of image quality metrics\footnote{Peak Signal to Noise Ratio (PSNR) measures the ratio between maximum possible power of signal and the mean squared error (differences) between the pixels of original and transformed image. Structural Similarity Index (SSIM) on the other hand takes into account structural differences by considering the luminance, contrast, and structure of the image.}, {\em viz.,} peak-signal-to-noise-ratio (PSNR) and structural similarity index measure (SSIM). It is also found that the rate of reduction in CNN's accuracy is lower in the proposed method than the simple reduction of ADC bits. The key contributions of this paper are:
\begin{itemize}
\item \textcolor{black}{An analog-domain 1D-HT-based compression method for machine vision applications that achieves the same classification accuracy as an 8-bits per pixel (BPP) image with as low as 3-BPP on ImageNet \cite{russakovsky2015imagenet} and CIFAR-10 datasets \cite{Krizhevsky2009}.}
\item \textcolor{black}{For larger images, {\em i.e.}, larger CIS array, the proposed method is further improved by discarding $50\%$ of the Hadamard transformed channels, which paves the way for reconfigurable image sensors for machine vision and data acquisition.}
\item \textcolor{black}{Embedded Hadamard Transform (EHT)-based pipelined ADC architecture to avoid additional circuitry for HT, reducing the power consumption and performance degradation due to thermal noise.}
\item Experimental validation of proposed idea with CIFAR-10 dataset. The CNN trained with full $8$-bit precision images can still identify the images whose pixels have been quantized with variable resolution to achieve low average BPP with an accuracy similar to original images. 
\end{itemize} 
\textcolor{black}{Image compression techniques are mostly implemented on digital images as previously mentioned. One previous work by \cite{kawahito1997cmos} on implementation of 8-point 2D-DCT reported a hardware intensive and power consuming circuit for achieving analog compression especially due to the presence of analog memory. In contrast, this paper proposes a 1D-HT which not only removes the need for analog memory but also is less prone to systematic errors as DCT is realized using irrational coefficients which are not easy to implement in hardware.}\\
\indent This paper is formulated as follows: Section \ref{sec: HT_and_res} describes the relation of HT with pixel resolution and the proposed idea. Section \ref{sec: Proposed Strategy and CNN performance} describes the image dataset used, CNN performance, and power consumption estimate of ADC for the proposed method as well as the {\em  baseline-quantization}. Section \ref{sec: HT1.5bckt} describes the proposed $1.5$-bit circuit incorporating HT and an analog gain. Section \ref{sec: MM} discusses the measurement setup for testing 100 images to study the effect of hardware induce errors on the CNN accuracy. Section \ref{sec: conclusion} is the conclusion drawn from the work along with its future scope. \textcolor{black}{Throughout this manuscript, the word ``pixel'' refers to the analog voltage that the CMOS image sensor in the CIS array produces, and ``{\em digitized pixel}'' or "{\em quantized pixel}" refers to the digitzed value.}\vspace{-0.5em}
\section{Proposed Strategy} 
\label{sec: HT_and_res}
\indent Efficient realization of orthogonal transforms in analog domain require that analog components like capacitors, resistors, current-sources, etc. should be able to efficiently scale as well as add/subtract analog signals. Unlike DCT, Haar, and Slant transforms which require that the signals be scaled by an irrational number like $\sqrt{2}$, HT on the other hand does not scale the analog signals as the HT matrix comprises of $\pm 1$. Thus, analog implementation of HT would require addition and subtraction\footnote{As signals in analog domain are typically differential, subtraction is easily realized by swapping the differential signals.} of analog signals which can be easily realized using capacitors. The HT transformed pixels would have different voltage dynamic range, facilitating {\em variable-resolution} quantization of the pixels using a Successive-Approximation Register ADC \cite{konwar2022deterministic, shah20178b} or pipelined ADC \cite{ravi2015speed} or Cyclic ADC \cite{kousik}. The power consumption of the ADC is a function of its resolution \cite{sundstrom2008power}. The relation between the ADC power consumption and the effective number of bits (ENOB) for a sampling frequency `$f_{S}$' is given by (\ref{eq:adc_power_bits}), where it is apparent that the power consumption of the ADC quadruples with $1$-bit increase in resolution.
\begin{equation}
    \label{eq:adc_power_bits}
    \textcolor{black}{P_{s} = 48KTf_{s}({2^{2ENOB}})}
\end{equation}
Here, `K' is the Boltzmann constant and `T' is the temperature. The proposed method, therefore, aims at reducing the average BPP consequently reducing ADC power without compromising the accuracy of the machine vision system.
\vspace{-1em}
\subsection{Relation between Variances of HT Image Pixels and Resolution of ADC}
\label{subsec: HTgain}
The HT is a linear, orthogonal transform that converts signals from the temporal or spatial domain to the frequency domain. It is used in various hardware applications due to the ease of efficient implementations \cite{9745740,hoshi2022real,dorrance2022energy,basiri2016efficient,4494572,1394536,1207040}, as the basis signals of the transform are $\pm 1$. For an $M$-point HT, the transform matrix, $H_{_M}$, is given by (\ref{eq:Hadamard_matrix}) \cite{horadam2012Hadamard}.
\begin{equation}
    H_{_M} = \frac{1}{\sqrt{2}} \begin{bmatrix}
        H_{_{M/2}} & H_{_{M/2}} \\
        H_{_{M/2}} & -H_{_{M/2}}
    \end{bmatrix}
    \label{eq:Hadamard_matrix}
\end{equation}
The identity matrix is defined as $H_1 = 1$. Excluding the scaling factor, the Hadamard matrix would always contain only $+1$ and $-1$, thus making it amenable for efficient hardware implementations\footnote{\textcolor{black}{Typically, for images, a two-dimensional transform is used. However, a 2D transform is more complex to implement in the analog domain, so a one-dimensional Hadamard transform is considered in this work.}}. \textcolor{black}{Applying an $M$-point 1D-Hadamard transform on an $I \times I$ image, where $I > M$, would result in a transformed image $Y$, given by (\ref{eq:Hadamard_transform_image}).}
\begin{equation}
    \label{eq:Hadamard_transform_image}
    \color{black}Y_i = [X_i^0 H_M, X_i^1 H_M, \dots, X_i^{k-1} H_M]
\end{equation}
\noindent where, \textcolor{black}{$X_i^0, X_i^1, \dots, X_i^{k-1}$ are $k$ segments of size $1 \times M$ in the $i^{\text{th}}$ row of the image}, where $k = \lceil I/M \rceil$. In the output image $Y$, each element of the first column would contain the corresponding sum of $M$ elements. The elements in the second column would contain alternating differences of $M$ elements, and so on. This pattern repeats with a periodicity of $M$.

\begin{figure}[htbp]
    \centering
    \vspace{-1em}
    \includegraphics[scale=0.5]{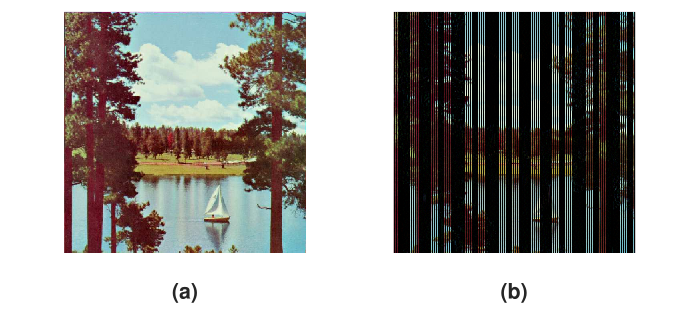} \vspace{-1em}
    \caption{\textcolor{black}{(a) Original image (8-BPP) and (b) 4-point 1D-HT image (1D-HT in digital domain, not in analog domain).}} \vspace{-1em}
    \label{fig:sample_Hadamard_transform}
\end{figure}

\begin{figure}[htbp]
    \centering
    \includegraphics[scale=0.3]{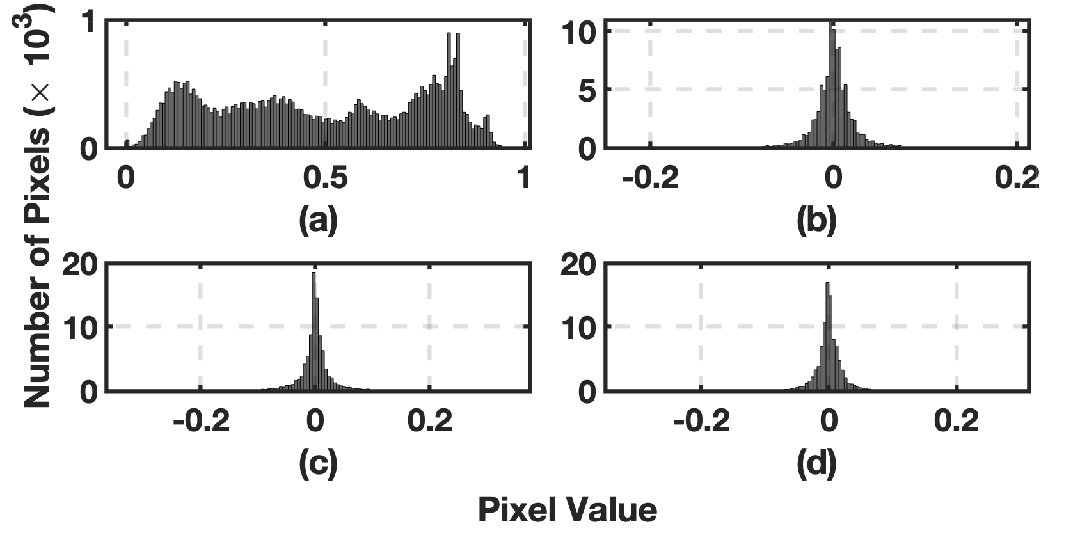}
    \caption{Histogram of the row sets (a) $Y_{4i}$, (b) $Y_{4i+1}$, (c) $Y_{4i+2}$, and (d) $Y_{4i+3}$ \textcolor{black}{of Fig. \ref{fig:sample_Hadamard_transform}(b).}}
    \label{fig:sailboat_rows_histogram}
    \vspace{-1em}
\end{figure}

A $4$-point HT is performed on a sample $512 \times 512$ color image to analyze the properties of the columns of the transformed image. Fig. \ref{fig:sample_Hadamard_transform} shows the original image (8-BPP image) and its 1D-HT counterpart obtained using (\ref{eq:Hadamard_transform_image}). 
There are four distinct sets of columns obtained after HT is performed on the pixels, indexed by $Y_{4i}, Y_{4i+1}, Y_{4i+2}, \text{and }Y_{4i+3}$, \textcolor{black}{where $i=0,1,2,\ldots (n/4-1)$ for an image of width $n$}. $Y_{4i}$ captures the DC component of the image as it is the sum of the pixel values. $Y_{4i+1}, Y_{4i+2}, \text{and }Y_{4i+3}$ captures the varying components of the image. Fig. \ref{fig:sailboat_rows_histogram} shows the histogram of the pixel values in the four sets of columns. The image pixels are normalized to $[0, 1]$ range before computing the Hadamard transform.

As shown in Fig. \ref{fig:sailboat_rows_histogram}, $Y_{4i}$ has a larger spread than the other three channels. The standard deviation ($\sigma$) of $Y_{4i}$ is the highest at $0.26$ while the standard deviation of the other three channels is $0.023, 0.039, \text{and }0.024$, respectively. This is because $Y_{4i}$ contains the sum of all the pixel values in the image, and images are generally slow-changing in a small $4 \times 4$ neighborhood.
The transformed pixels of $Y_{4i}$ spread across the complete full-scale range normalized to 1 having the highest $\sigma$. The $\sigma$ of other three channels are much lower and thus the ratio of the $\sigma$ of the channels to the first channel gives the values of gains that need to be multiplied. Since now, the last three channels are amplified to the full scale, they can be quantized to the lower number of bits. If the quantization bits required for digitizing the first channel $Y_{4i}$ is $N_0$, then the number of bits for digitizing the rest of the channels is,\vspace{-1em}
 \begin{equation}
  \label{eq: sigma}
    N_{j}=N_{0}-\alpha_{j}, \vspace{-0.6em}
\end{equation}
where, $\textcolor{black}{\alpha_{j}=\text{floor}\left(\log_{2}(\sigma_{1}/\sigma_{j})\right)}$. In (\ref{eq: sigma}), $\sigma_{1}$ refers to the $\sigma$ of channel $1$, while the $\sigma$ of channel-$2$, channel-$3$, and channel-$4$ is given by $\sigma_j$ for $j = 2,\ 3,\ \text{and } 4$. 

\begin{figure}
\centering
\includegraphics[scale=0.25]{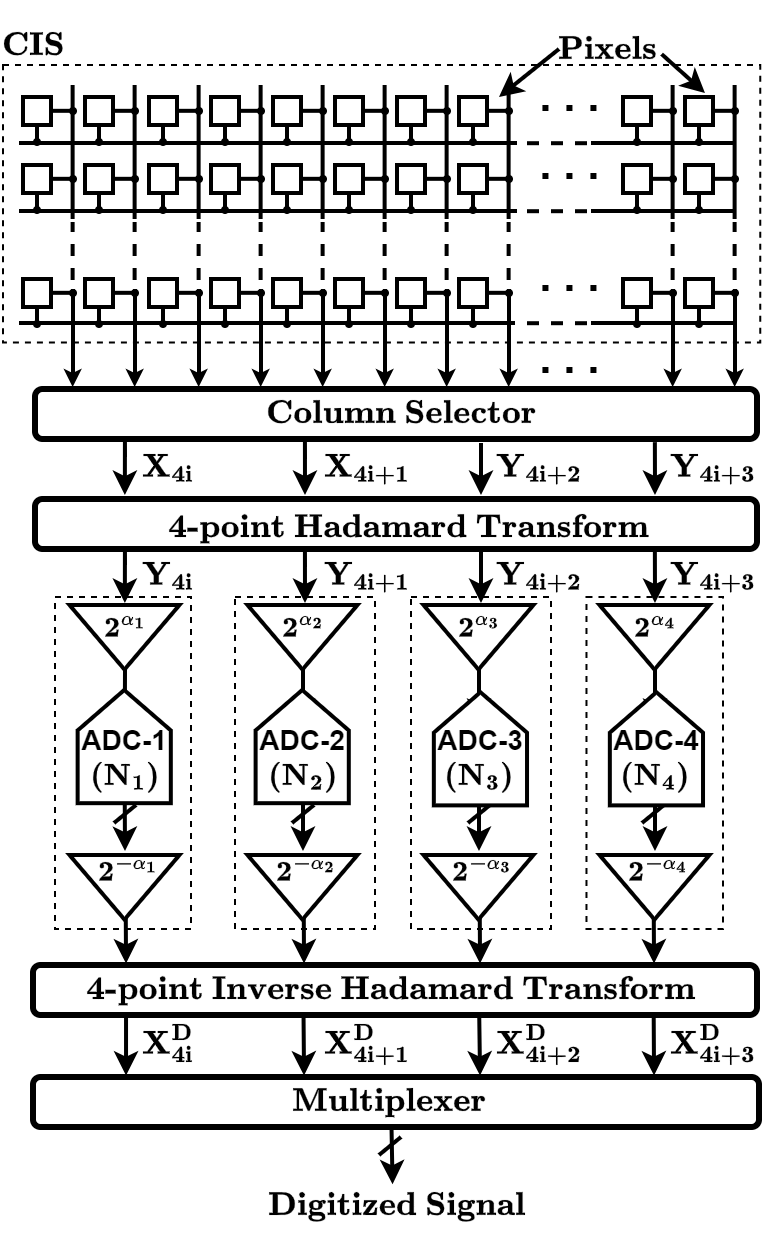}
\vspace{-1em}
\caption{\textcolor{black}{Block diagram for modeling the proposed method. For example, in Fig. \ref{fig:sailboat_rows_histogram}, the $\sigma$ of $Y_{4i}=0.2526$, $Y_{4i+1}=0.0232$, $Y_{4i+2}=0.0397$ and, $Y_{4i+3}=0.0239$. Hence, $\alpha_{2-4}=3$, implying that if the first channel $Y_{4i}$ is digitized to full scale resolution of $N_{1}=N_{0}=8$b ($\alpha_1=0$) then rest channels are digitized to $N_{2-4}=5$b after putting a gain of $8$.}}
\vspace{-1em}
\label{fig: BD_new}
\end{figure}

Fig. \ref{fig: BD_new} shows the block diagram of the proposed {\em variable-quantization} ADC consisting of four parallel ADCs. The sampling rate $f_s$ of the ADC is set in a manner that guarantees the frame rate of the CIS is properly maintained. Each channel of this ADC performs signal transformation followed by variable pixel quantization such that ADC-$1$ digitizes `$N_{1}$' bits, ADC-$2$ digitizes `$N_{2}$' bits, ADC-$3$ digitizes `$N_{3}$' bits and ADC-$4$ digitizes `$N_{4}$' bits. Similarly, the ADC architecture for {\em baseline-quantization} also consists of four parallel ADCs, each of them digitizing the full scale resolution of `$N_{0}$' bits.
The analog signal from the pixels undergo 4-point HT thus giving four output channels to be digitized (each channel is digitized with different resolution as the dynamic range of the analog signals is different in each channel as mentioned earlier). As the analog signals in each channel has different variance as mentioned earlier, the signals are scaled by $2^{\alpha_j}$ resulting in the quantizer following the gain stage having a resolution given by (\ref{eq: sigma}). Inverse Hadamard is then performed along with the respective reverse gains on this digitized signal per channel. The signal thus obtained from the four channels is then combined to get the final digital output of the original image. Thus, each of the channel takes four samples per conversion cycle and perform the required arithmetic operation corresponding the HT-matrix\footnote{As HT results in different pixels getting quantized with different resolutions, the overall image can be considered to be quantized with an average bit-per-pixel which is much lower than the original. Thus, throughout the paper as the analysis uses the HT images with pixels digitized with variable resolution which result in average BPP being lower, we refer to such processing as processing on {\em transformed images}.}.

\section{Validation of the Proposed Method for Machine Vision Application}
\label{sec: Proposed Strategy and CNN performance}
The analog gains ($2^{\alpha_{j}}$) and the corresponding resolution of the ADC given by (\ref{eq: sigma}) will differ according to the incoming image signal, {\em i.e.}, CIS output. The dynamic adjustment of resolution of ADC can be cumbersome and it is clear from Fig. \ref{fig: BD_new} that there is a requirement to fix the bits per channel (BPC) of the ADC. This will result in some degradation in the digitized image signals and therefore there is a need to perform a detailed analysis to propose a model that will work efficiently for a wide range of images\footnote{Throughout the text 4-point HT is used to obtain transformed image.}. This section explains
 \begin{itemize}
    \item the procedure to validate the idea and obtain fixed BPC,
    \item the performance of the proposed method with {\em baseline-quantized} images from various datasets like CIFAR10 and ImageNet,
    \item the performance of CNN on $5000$ transformed images from ImageNet dataset built on a Pre-trained ResNet architecture for varying image sizes,
    \item the accuracy of a second CNN architecture based on VGG style network trained on 8-bit CIFAR-10 images when tested with transformed images. 
 \end{itemize}
\vspace{-1em}
\subsection{System-level Validation on $200$ Images to Obtain BPC}
\label{sec:cifar_validation}
 Fig. \ref{fig: BD_new} is modeled in MATLAB such that it stores the gains needed per channel based on the ratio of $\sigma$ \emph{w.r.t} the first channel for a set of $200$ 4-point HT images. The average values of the $200$ gains obtained for the four channels are taken as the final values. \textcolor{black}{It is to be noted that since the images vary in dimensions for ImageNet dataset, the gains are obtained by clipping the images to the nearest dimension divisible by 4 so as to enable the application of HT}. For the ImageNet training set, gains $1, 8, 4, 8$ are obtained while for CIFAR-$10$ training set, the gains are $1, 4, 2, 2$, based on the ratios of the $\sigma$'s. However, for CIFAR-10, similar performance of {\em image-metrics}, {\em i.e.}, PSNR and SSIM \cite{1284395}, is obtained with gains $1, 8, 4, 8$ as well. Therefore, based on the {\em image-metrics} of $200$ random images from both datasets, gains of $1, 8, 4, 8$ are fixed for the $4$-channels, resulting in resolutions of $N_1= N_0, N_2= N_0-3, N_3= N_0-2, N_4= N_0-3$ for a full-scale resolution of $N_0$ bits. Considering $N_{0}=8$, the values of $N_{1\ldots 4}$ will be $8, 5, 6, 5$ in the block diagram of Fig. \ref{fig: BD_new}, resulting in an average BPP of $6$-bits. To compare this method with the {\em  baseline-quantization} method, an analysis on $100$ random images from ImageNet test set scaled to different pre-decided image sizes and BPP is done. The BPP is reduced by varying $N_{0}$ from $8$b to $4$b thereby resulting in the average BPP for 4-point HT images varying from $6$-BPP to $2$-BPP. 
\begin{figure}
\centering
\includegraphics[scale=0.22]{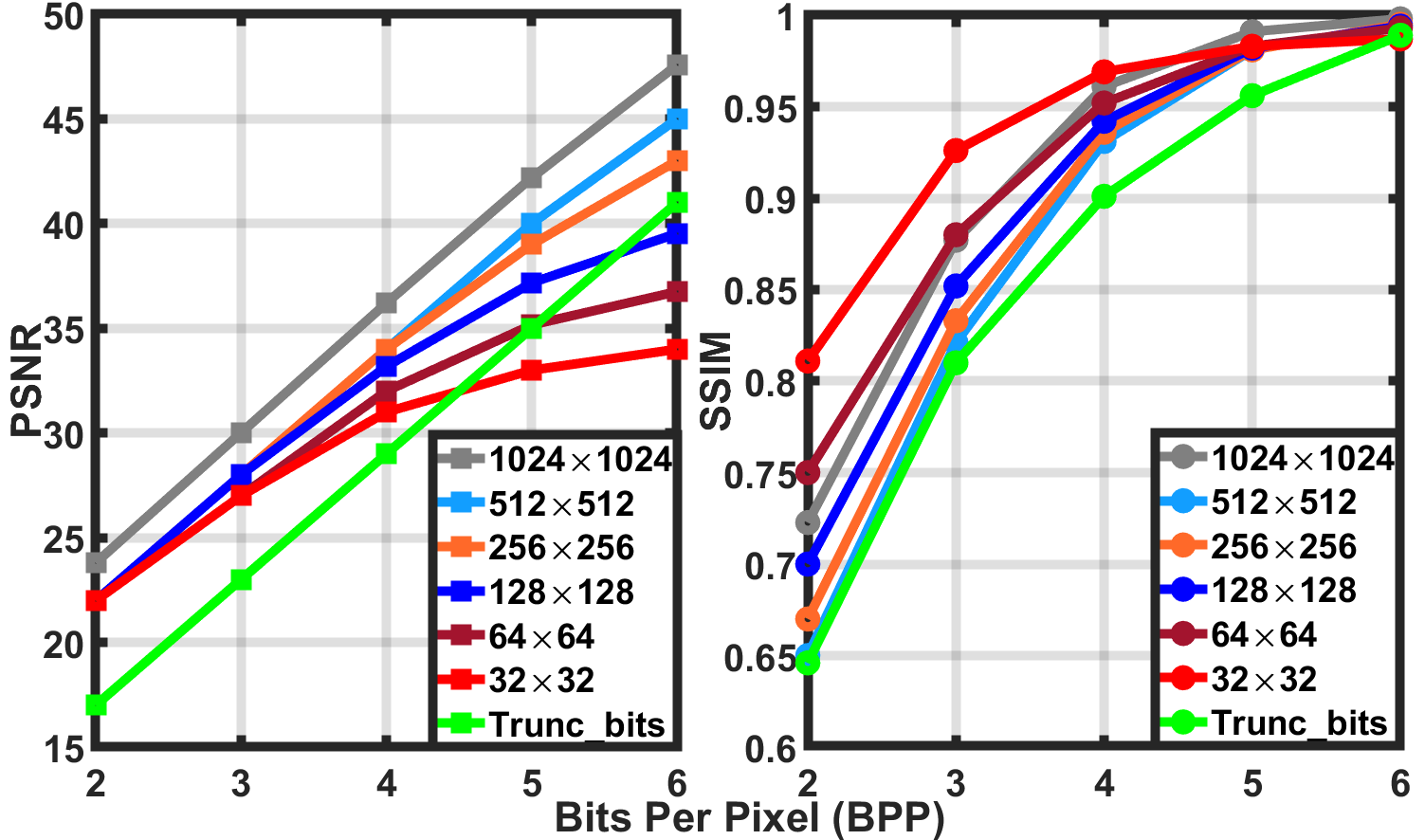}
\caption{\textcolor{black}{\small{Variation of Image metrics \emph{i.e.} PSNR and SSIM \emph{w.r.t} BPP for different image sizes for proposed method and {\em baseline-quantization} method. The {\em baseline-quantized} image metrics (Trunc\_bits\_PSNR and Trunc\_bits\_SSIM) \emph{w.r.t} varying BPP remain same across different image dimensions, therefore denoted by single curves}}}
\vspace{-2em}
\label{fig: PSNRandSSIM}
\end{figure}

\indent Fig. \ref{fig: PSNRandSSIM} compares the {\em image-metrics} for different BPP for images of various sizes and documents the behavior of the {\em  baseline-quantization} method and the proposed idea for all the cases. 

\indent Approximately, $100$ random images from ImageNet are resized\footnote{The resizing is done using Bi-linear interpolation where the value of new pixel is calculated based on the weighted average of the four nearest pixels in the original image.} to $(I\times I)$, with I$= 2^{x}$ where $x = 5$ to $10$. These images are then processed using the proposed method and the {\em  baseline-quantization} method. The comparison is made between these two methods based on the {\em image-metrics} of the processed images compared to the original. The PSNR and SSIM for different image sizes with pixels quantized using the {\em  baseline-quantization} method is denoted by Trunc\_bits\_PSNR and Trunc\_bits\_SSIM in Fig. \ref{fig: PSNRandSSIM}. It is worth noting that the PSNR and SSIM degradation with reduced BPP is independent of the image size (does not matter if the image size is $32 \times 32$ or $1024\times 1024$) when the image pixels are quantized using the {\em baseline-quantization} method. However, the {\em image-metrics} with varying BPP differ according to the image size using the proposed method (ref. Fig. \ref{fig: PSNRandSSIM}). The BPP for the proposed method is calculated as follows: if the image pixels are all quantized based on (\ref{eq: sigma}) for $N_{0}=8$ and the resulting resolution in the 4-channels are $8, 5, 6, 5$, then average BPP is $(8+5+6+5)/4=6$.

Thus, from Fig. \ref{fig: PSNRandSSIM} it can be seen that for all the image sizes considered, the PSNR for the proposed method is higher than that of the {\em  baseline-quantization} method for $BPP<5$ with the exception of $32\times32$ images where the PSNR lowers for $BPP>4.5$. SSIM is always on the higher side for the proposed method with significant difference starting at $BPP>3.5$. Notably, for image sizes larger than $256\times 256$, the {\em image-metrics} of the proposed $N$-bit system matches that of the $(N+1)$-bit {\em  baseline-quantization} method, clearly showing a 1-bit improvement. Thus, the proposed method using 1D-HT is beneficial for large image sizes across all BPP values whereas for smaller images, it proves to be better for low BPP. Thus, the proposed method provides better similarity of the proposed images to the original as compared to the {\em  baseline-quantization} method.
\vspace{-1em}
\subsection{ADC Resolution and Power Consumption Trends}
\label{sec:power_compare}
This section provides an analysis of ADC power dissipation as a function of ADC resolution for two types of ADCs, \emph{viz.} Pipelined ADC \cite{cho,sarma2017250} and Successive Approximation Register (SAR) ADC \cite{shah20178b,konwar2022deterministic}. The study conducted by Sundström {\em et al.} in \cite{sundstrom2008power} offers an in-depth examination of ADC power consumption, serving as a foundation for the subsequent analysis in this section. The insights derived from this analysis are then applied in Section \ref{subsec: ResNet50_performance} to characterize the power requirements of the ADC for the proposed method as well as the {\em  baseline-quantization} method. The power consumption of the ADC quadruples with each increase in ENOB, \textcolor{black}{$N$}, as shown in (\ref{eq:adc_power_bits}). The analysis presented in \cite{sundstrom2008power} uses the expression in (\ref{eq:adc_power_bits}) to define the power consumption of a noise limited pipelined ADC realized using $1.5$-bit stages as given by (\ref{eq:pnoise}).
\begin{align}
         \label{eq:pnoise}
            P_{pn} = 9\left(1+2N\frac{V_{eff}}{V_{FS}}\log_{e}{2}\right)P_{s}
   \end{align}
In  (\ref{eq:pnoise}), $V_{eff}$ is taken as $(V_{GS}-V_{TH})/2$ for a MOS transistor in strong inversion where `$V_{GS}$' is the gate-source voltage and `$V_{TH}$' is the threshold voltage of the MOSFET. Denoting the input capacitance of a minimum-sized inverter as {\em $C_{min}$}, where, $C_{min}$ depends on the technology node,`$N$' as the resolution of ADC, `$f_{S}$' as sampling frequency and `$V_{FS}$' as the {\em full scale range}\footnote{The ADC full scale range is considered to be $0$ to $V_{ref}$.} of the ADC, the expression for total power consumed by a pipelined ADC limited by process parameters and thermal noise is given by (\ref{eq:pipe}),
\begin{equation}
         \label{eq:pipe}
            P_{pipe} = P_{pn}+2NC_{min}V^{2}_{FS}f_{s}\left(1+6N\frac{V_{eff}}{V_{FS}}\log_{e}{2}\right)
   \end{equation}
Thus, based on (\ref{eq:pipe}) the power consumption of an $8$-bit pipelined ADC in $90$-nm CMOS process for a sample-rate of $f_{S}=50 MHz$, $V_{ref}=1V$, $V_{eff}=0.1V$ , $C_{min}=1 fF$, $K=1.38\times 10^{-23}$, and $T=300 K$ is $15.82 \mu$W.\\
\indent The power consumption of an $N$-bit SAR ADC with a unit DAC capacitance $C_{unit}$ and reference voltage $V_{ref}$ is given by (\ref{eq:SAR})\cite{5437496},\\
\vspace{-1em}
\begin{equation}
         \label{eq:SAR}
           \textcolor{black}{ P_{sar}= f_{s}\sum_{i=1}^{N-1} (2^{N-2-i})C_{unit}V^{2}_{ref}}
   \end{equation}
which results in a power consumption of $15.24 \ \mu$W for an $8$-bit SAR ADC operating at a sample-rate of $f_{S}=50 MHz$ and having $V_{ref}=1V$ and $C_{unit}=4.8 fF$. Table \ref{tab:simulated_power_results} shows the normalized power for different ADC resolution, $N$. These calculations are done for a single channel\footnote{All power numbers quoted in the subsequent tables are power per channel. For RGB, the reported numbers must be multiplied by a factor of $3$.}. \vspace{-1em}
\begin{table}[htbp]
    \renewcommand{\arraystretch}{1.1}
    \centering
    \caption{Normalized power consumption for SAR ADC and Pipelined ADC \emph{w.r.t} 8-BPP for different ADC bit precision.}
    \label{tab:simulated_power_results}
    \begin{tabular}{|c|c|c|c|c|c|c|c|}
    \hline
    \Centerstack{\textbf{Power}} & \multicolumn{7}{c|}{\textbf{ADC bit precision (N)}} \\
     \cline{2-8}
     \Centerstack{\textbf{(norm.)}}& \textbf{8} & \textbf{7} & \textbf{6} & \textbf{5} & \textbf{4} & \textbf{3} & \textbf{2} \\
    \hline
    $P_{pipe}$ & $1$ & $0.355$ & $0.175$ & $0.107$ & $0.069$ & $0.043$ & $0.023$ \\
    \hline
    $P_{sar}$ & $1$ & $0.495$ & $0.243$ & $0.117$ & $0.055$ & $0.024$ & $0.0079$ \\
    \hline
    \end{tabular}
\end{table} \vspace{-1em}
\begin{figure*}[htbp]
\centering
\includegraphics[scale=0.28]{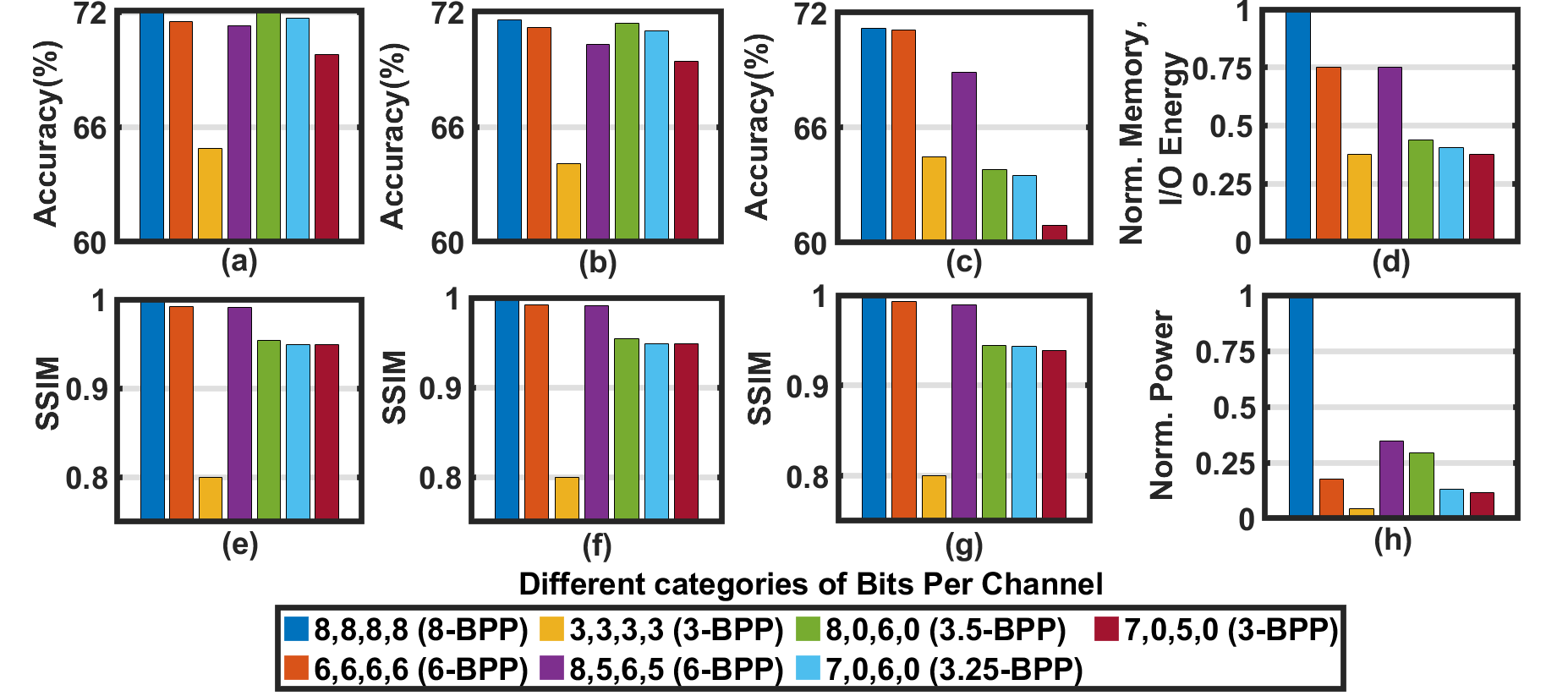} \vspace{-1em}
\caption{\textcolor{black}{Bar chart showing (a) accuracy of CNN for $1024\times 1024$ image (b) accuracy of CNN for $512\times 512$ image, (c) accuracy of CNN for $256\times 256$ image, (d) Memory required and I/O Energy consumed per bit, normalized \emph{w.r.t} $8$-BPP (e) Mean SSIM of $1024\times 1024$ images (f) Mean SSIM of $512\times 512$ images, (g) Mean SSIM of $256\times 256$ images, (h) ADC power consumption per bit, normalized \emph{w.r.t} $8$-BPP for the {\em proposed} and {\em baseline-quantization} methods for different {\em BPCs}.\\
\textbf{Note:} The combinations (8888, 6666, 3333) indicate {\em baseline-quantization}; rest of the BPCs are for {\em proposed-method.} Except for the CNN accuracy, all other parameter trends are independent of the image size.}} \vspace{-0.5em}
\vspace{-1.5em}
\label{fig: Acc_results}
\end{figure*}

Thus, (\ref{eq:adc_power_bits})-(\ref{eq:SAR}) can be used to find the normalized power consumption of the variable-resolution ADCs used to digitize the 4-channels of the 4-point HT output. The power consumption due to the presence of HT circuit is not considered as the paper proposes (ref. Section \ref{sec: HT1.5bckt}) a pipelined ADC architecture that performs $1$-bit digitization while performing HT in the $1^{st}$-stage. Based on Table \ref{tab:simulated_power_results} the normalized power consumption of $4$-channel ADC having resolutions 8-bit, 5-bit, 6-bit, and 5-bit is given by (\ref{eq:ptotal}), thereby translating to normalized-power/channel given by (\ref{eq:ptotal1}). 
\begin{align}
         \label{eq:ptotal}
            P_{4\text{-}ch}^{total} = (1+0.107+0.175+0.107) = 1.389\\
         \label{eq:ptotal1}
            P_{pc} = P_{4-ch}^{total}/4 = 0.34725.
\end{align}
\indent Besides the ADC power consumption, the power consumption for transmitting as well as storing the digitized pixel values need to be considered. Both the energy required to transmit as well as store a pixel digitized to $N$-bits is proportional to $N$. As the proposed technique results in pixels quantized using variable resolution without compromising the image quality, the resultant BPP reduces, thereby reducing the energy required to transmit as well as store the \textcolor{black}{digitized} pixel. A comparison of the {\em proposed method} with the {\em baseline-quantization} is done in the following section on the basis of these parameters.
\vspace{-1em}
\subsection{Performance of ResNet Model with transformed Images (ImageNet Dataset)}
\label{subsec: ResNet50_performance}
\textcolor{black}{Image classification is used as the target machine vision task to evaluate the proposed method. The performance metric of interest is the classification accuracy of the machine learning model on a set of test images. In this work,} a ResNet50 model pre-trained on the ImageNet dataset is chosen \cite{tensorflow2015-whitepaper}. An evaluation set consisting of $5000$ images is prepared by randomly sampling $5$ images from each of the $1000$ classes in the ImageNet validation set. \textcolor{black}{The top-1 accuracy of the CNN calculated for the $5000$ images from this evaluation set is $69.1\%$}. To observe the effect of image size on the performance of the proposed method, the evaluation set images are resized to various spatial resolutions, \text{viz.}, $1024\times1024$, $512\times512$, and $256\times256$. These images are then processed using the proposed method and {\em  baseline-quantization}, respectively. The processed images are resized back to $224\times224$ before sending to the pre-trained CNN model for inference due to the input layer size constraint of the CNN. It is to be noted that the pre-trained model is used without any transfer learning or fine-tuning. 

\textcolor{black}{Fig. \ref{fig: Acc_results} shows the bar chart indicating the {\em performance comparison}\footnote{The size of the images is not considered in calculation as it does not have an effect on the performance trend of the parameters.} of the {\em proposed} method {\em w.r.t.} the {\em baseline-quantization} on the basis of CNN accuracy, mean SSIM, ADC power consumption, transmit energy consumed per bit, and storage-energy per bit. This is done to bring out the optimal ADC architecture that gives a trade-off between accuracy and power. As mentioned previously, except for the CNN accuracy and SSIM, all other parameters are normalized \emph{w.r.t} pixels quantized to $8$-BPP.\\
Following are the three broad categories considered to plot Fig. \ref{fig: Acc_results} for achieving an optimal low power ADC architecture:
\begin{enumerate}  
\item \textbf{Truncation of Bits} ($8888$, $6666$, $3333$): Considered as {\em baseline} for comparison, this is the best method to follow if $6$-BPP quantization is desired as it is optimal in terms of high CNN accuracy at a lower power consumption. The {\em baseline} 8-BPP give the highest CNN accuracy which comes with the penalty of high power consumption. While the power of {\em baseline} $3$-BPP images is the lowest, the CNN accuracy and SSIM falls considerably making them unfit for edge applications.
\item \textbf{Non-uniform quantization} ($8565$): The performance of non-uniform quantization achieved using the four channel architecture is shown for $N_{0}=8$. This {\em proposed} $6-$BPP model gives performance on par with the {\em baseline} $6-$BPP images but consumes more power thus making it unsuitable for larger images.
\item \textbf{Elimination of ADC channels} ($8060$, $7060$, $7050$): The two channel ADC architectures obtained by elimination of two of the channels from the four channel ADC result in a CNN accuracy approximately equal to {\em baseline} $8-$BPP and $6-$BPP images along with very low power consumption, I/O energy and memory requirement for $1024\times1024$ and $512\times512$ images. Although for $256\times256$ images, the {\em baseline} $3$-BPP images show slightly better CNN accuracy than the {\em proposed} $3.5$ BPP images, its SSIM is extremely low ($\approx 0.8$) making it unfit for data collection purposes. The SSIM for the proposed architectures are greater than $0.94$ making them fit for both inference and data collection for further processing by other applications.
\end{enumerate}
Therefore, the two channel architectures are optimal for larger images while the proposed four channel architecture works better for smaller images as will be seen in the subsequent sections.}
\vspace{-1em}
\subsection{Performance of CNN with transformed Images (CIFAR-10 dataset)}
\label{sub: CNN_performance}
\indent As the drop in accuracy seems to be more with the reduction of image size \textcolor{black}{(Fig. \ref{fig: Acc_results})}, further analysis is done on the CIFAR-10 dataset \cite{Krizhevsky2009}, which has extremely small $32\times32$ images. Prior works have analyzed the accuracy of CNNs when trained with low bit-resolution images \cite{Sanjeet2023}. The 7 convolutional layer VGG-styled network described in \cite{Sanjeet2023} is used in this work, and is detailed in Table \ref{tab:cifar_cnn}. The network achieves a classification accuracy of $91.01\%$ on the test set when trained with the original 8-bit images. This network is considered the {\em  baseline-network} in this section. It is shown in \cite{Sanjeet2023} that the network can be trained to classify $3$-BPP truncated images with a $\approx 2\%$ drop in accuracy.

\begin{table}[h]
    \renewcommand{\arraystretch}{1.1}
    \centering
    \caption{Architecture of the VGG-style network used.}
    \label{tab:cifar_cnn}
    \begin{tabular}{|c|c|c|c|}
    \hline
    \textbf{Layer} & \textbf{Kernel Size} & \textbf{Output Size} & \textbf{Number of Parameters} \\
    \hline
    Conv 1 & $3 \times 3$ & $32 \times 32 \times 32$ & $864$ \\
    \hline
    Conv 2 & $3 \times 3$ & $32 \times 32 \times 64$ & $18,432$ \\
    \hline
    Maxpool & $2 \times 2$ & $16 \times 16 \times 64$ & - \\
    \hline
    Conv 3 & $3 \times 3$ & $16 \times 16 \times 64$ & $36,864$ \\
    \hline
    Conv 4 & $3 \times 3$ & $16 \times 16 \times 64$ & $36,864$ \\
    \hline
    Maxpool & $2 \times 2$ & $8 \times 8 \times 64$ & - \\
    \hline
    Conv 5 & $3 \times 3$ & $8 \times 8 \times 128$ & $73,728$ \\
    \hline
    Conv 6 & $3 \times 3$ & $8 \times 8 \times 128$ & $147,456$ \\
    \hline
    Conv 7 & $3 \times 3$ & $8 \times 8 \times 128$ & $147,456$ \\
    \hline
    Maxpool & $2 \times 2$ & $4 \times 4 \times 128$ & - \\
    \hline
    Avg. Pool & $2 \times 2$ & $2 \times 2 \times 128$ & - \\
    \hline
    Output & $512 \times 10$ & $10$ & $5130$ \\
    \hline
    \end{tabular}
    \vspace{-1em}
\end{table}

The {\em  baseline-network} is tested on images transformed using the proposed method described in Section \ref{sec: HT_and_res}, with lower resolutions in the last three ADC channels. Table \ref{tab:simulated_transformed_results} shows the accuracy of the network with varying resolution of the first channel to get different BPC. There is a significant drop in accuracy at low resolution levels, particularly when the BPP falls below 5.

One way of improving the performance of the CNN is by fine-tuning the network with transformed images from the training set. To validate this idea, a dataset is generated from the CIFAR-10 training set with all levels of resolutions mentioned in Table \ref{tab:simulated_transformed_results}. For each resolution level, \textcolor{black}{$10,000$ images from the training data are randomly sampled and transformed to generate the training data, resulting in a total training data size of $80,000$. Only the first two convolutional layers and the output layer are fine-tuned, implying all the remaining layers are not updated. Fine-tuning is done by training the network with a smaller learning rate on the newly generated transformed training set for $30$ epochs.} Table \ref{tab:simulated_transformed_results} displays the network's performance on the transformed test set after fine-tuning. \textcolor{black}{For each row of Table \ref{tab:simulated_transformed_results}, the accuracy is reported by transforming and quantizing the entire CIFAR-10 test set with the corresponding BPC. The fine-tuned network, however, is the same for all the test cases.} The network suffers from only $\approx 1.5\%$ drop in accuracy post-fine-tuning for BPP levels as low as $3$ as compared to original $8$-bit images' accuracy.

\begin{table}[htbp]
    \renewcommand{\arraystretch}{1.1}
    \centering
    \caption{Comparison of CNN Performance and Normalized Pipelined ADC power consumption for $32\times 32$ pixels following the proposed method}
    \label{tab:simulated_transformed_results}
    \begin{tabular}{|c|c|c|c|c|c|c|}
    \hline
     \textbf{} &  & \multicolumn{2}{c|}{\textbf{{Baseline-net.} CNN}} & \multicolumn{2}{c|}{\textbf{Fine-tuned CNN}} & \\
     \cline{3-6}
    \textbf{BPC} & \textbf{BPP} & \textbf{Acc. (\%)} & \textbf{Loss} & \textbf{Acc. (\%)} & \textbf{Loss} & $\mathbf{P_{pc}}$\\
    \hline
    Original & $8$ & $91.01$ & $0.394$ & $91.65$ & $0.379$ & $1$\\
    \hline
    $(8, 5, 6, 5)$ & $6$ & $90.00$ & $0.428$ & $91.39$ & $0.383$ & $0.35$\\
    \hline
    $(7, 4, 5, 4)$ & $5$ & $89.09$ & $0.453$ & $91.38$ & $0.385$ & $0.15$\\
    \hline
    $(6, 3, 4, 3)$ & $4$ & $85.47$ & $0.579$ & $91.15$ & $0.392$ & $0.08$\\
    \hline
    $(8, 0, 6, 0)$ & $3.5$ & $79.00$ & $0.750$ & $90.65$ & $0.415$ & $0.29$\\
    \hline
    $(7, 0, 6, 0)$ & $3.25$ & $78.71$ & $0.880$ & $90.44$ & $0.419$ & $0.13$\\
    \hline
    $(7, 0, 5, 0)$ & $3$ & $73.50$ & $1.050$ & $90.15$ & $0.425$ & $0.12$\\
    \hline
    $(5, 2, 3, 2)$ & $3$ & $72.90$ & $1.121$ & $90.07$ & $0.430$ & $0.05$\\
    \hline
    $(4, 1, 2, 1)$ & $2$ & $47.09$ & $2.473$ & $79.68$ & $0.793$ & $0.03$\\
    \hline
    \end{tabular}\vspace{-1em}
\end{table}
For CIFAR-10 images, the accuracy shows a good consistency till $BPP=4$ and reduces by only $0.5\%$, with the ADC power showing a remarkable reduction of $\approx92\%$ as compared to original 8-bit images. \textcolor{black}{In fact} for proposed $3$-BPP, the accuracy still remains at $\approx 90\%$, which is very close to the $8$-BPP digitized image.

\begin{table}[h]
    \renewcommand{\arraystretch}{1.1}
    \caption{Comparison of the proposed method with prior works.}
    \label{tab:comparison_table}
    \centering
    \begin{tabular}{|c|c|c|c|}
        \hline
        \textbf{Work} & \textbf{BPP} & \textbf{Network} & \textbf{Accuracy (\%)} \\
        \hline
        \cite{kisku2023} & 4 & VGG16 & 59.45 \\
        \hline
        \cite{kisku2023} & 2 & VGG16 & 41.94 \\
        \hline
        \cite{Sanjeet2023}$\color{black}^{\S}$ & 3 & VGG7 & 87.02 \\
        \hline
        \cite{Guo2018} & 5 & BNN* & 88.81 \\
        \hline
        \cite{Zhang2021}$\color{black}^{\S}$ & 5 & ResNet-20 BNN* & 89.10 \\
        \hline
        \textbf{This work$^\dagger$$\color{black}^{\S}$} & \textbf{3.5} & \textbf{VGG7} & \textbf{90.65} \\
        \hline
        \textbf{This work$^\ddagger$$\color{black}^{\S}$} & \textbf{3} & \textbf{VGG7} & \textbf{90.15} \\
        \hline
        \multicolumn{4}{l}{*BNN = Binary Neural Network.}\\
        \multicolumn{4}{l}{$^\dagger$ For BPC of (8,0,6,0) as shown in Table \ref{tab:simulated_transformed_results}.}\\
        \multicolumn{4}{l}{$^\ddagger$ For BPC of (7,0,5,0) as shown in Table \ref{tab:simulated_transformed_results}.}\\
        \multicolumn{4}{l}{$\color{black}^{\S}$ \textcolor{black}{The training data contains reduced precision images.}}
        \vspace{-1.5em}
    \end{tabular}
\end{table}
Table \ref{tab:comparison_table} compares the bits per pixel and accuracy of the proposed work with prior works. As shown in Table \ref{tab:comparison_table}, the proposed work outperforms prior works in terms of accuracy using only $3$-BPP. This shows that the proposed method of transformation is a good fit for machine vision applications, as the drop in image quality due to aggressive bit-reduction (low resolution) can be recovered in the neural network by appropriate fine-tuning. This claim is further strengthened with measurement results in the subsequent sections.

\section{Proposed HT embedded ADC architecture with Circuit and System Level Validation}
\label{sec: HT1.5bckt}
This section describes the proposed HT circuit embedded in the $1^{st}$ stage of the pipelined ADC to avoid thermal-noise and power penalty of a dedicated HT circuit preceding the ADC. The HT circuit is embedded in the popular $1.5$-bit non-flip around topology such that it performs HT and $1.5$-bit digitization in the same cycle. System level modeling in MATLAB as well as circuit level simulations in Cadence are done for the proposed architecture to validate its robustness in presence of non-idealities like capacitor mismatch and finite \textcolor{black}{op amp} gain. \vspace{-1em}
\subsection{Switched Capacitor Circuit based $1.5$-bit Stage of ADC}
\label{sub: SCckt}
\begin{figure}[htbp]
\centering
\includegraphics[scale=0.125]{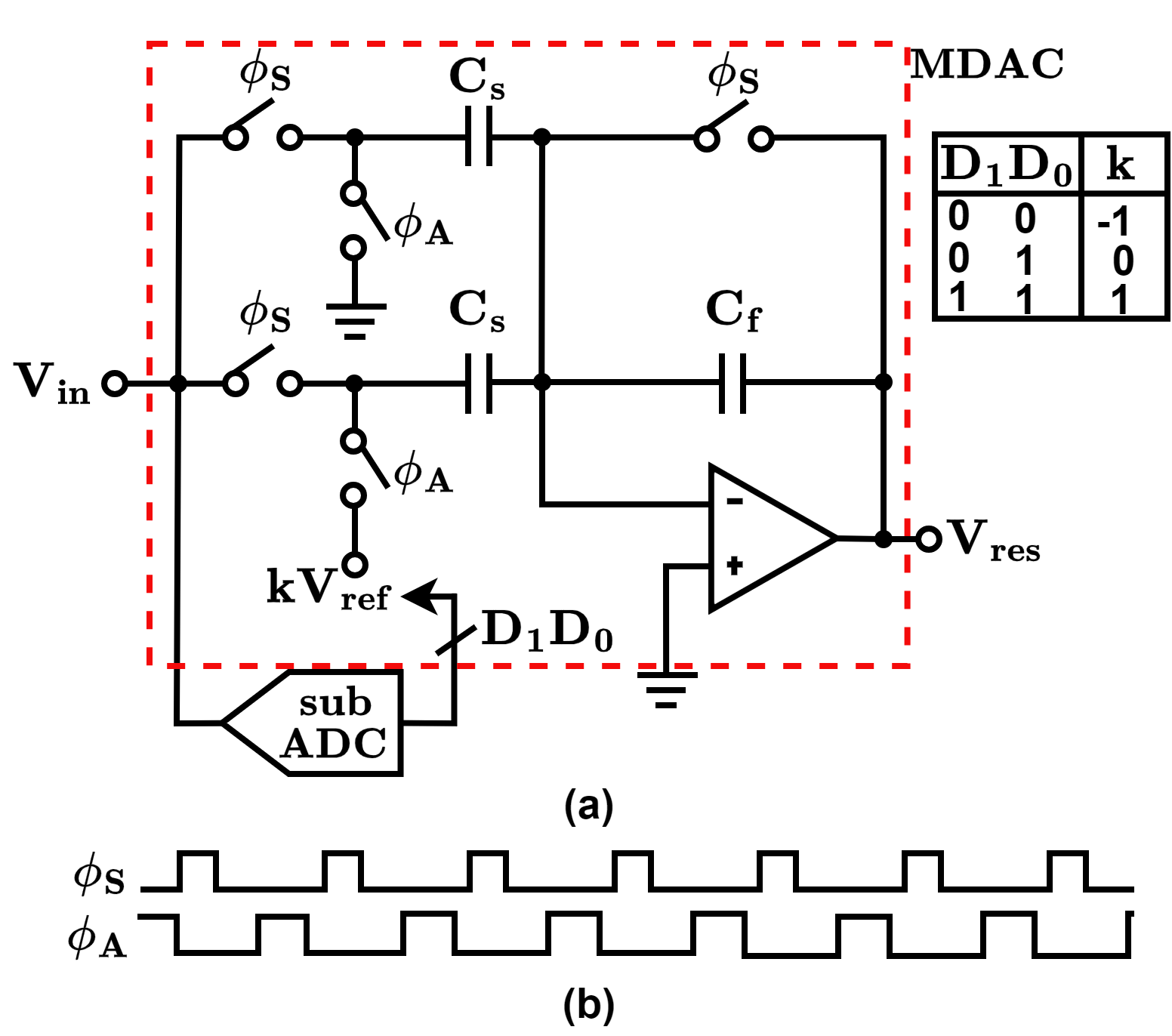}
\caption{(a) Architecture for a $1.5$-bit ADC (b) clocks needed for operation of $1.5$-bit ADC.} \vspace{-1em}
\label{fig: 1.5bADC}
\end{figure}

\indent Earlier works have shown \textcolor{black}{Switched Capacitor} (SC) implementation of various transforms like DCT, HT \emph{etc.} \cite{brandao2005switched,rahimi2018energy,mal2004analog,kawahito1997cmos}. Fig. \ref{fig: 1.5bADC} shows a $1.5$-bit SC stage of a pipelined ADC, that has a sub-ADC comprising of two comparators and a \textcolor{black}{Multiplying Digital-to-Analog} converter (MDAC) for generating the residue voltage, $V_{res}$,  which is the difference between the input voltage and analog equivalent of digitized output from the sub-ADC \cite{cho,sarma2017250}. 
When clock $\phi_{S}$ is high, the input voltage $V_{in}$ gets sampled on the capacitors $C_s$ and the \textcolor{black}{op amp} is reset. The total sampled charge is given by $Q_{s} = 2V_{in}C_{s}$. When $\phi_{A}$ is high, the charge is transferred to $C_{f}$ with the reference voltage ($\pm V_{ref}$) or $0$, connected to one of the sampling capacitors based on the sub-ADC output, while the other sampling capacitor is connected to ground.  Thus, at $\phi_{A}$ the total charge in all the capacitors is, $Q_{h}=V_{res}C_{f}+kV_{ref}C_{s}$. Considering $C_{s}=C_{f}$ and using the principle of conservation of charge where, $Q_{s}=Q_{h}$ the residue voltage is given by (\ref{eq:res1p5}). This residue voltage is digitized by the subsequent stages of the pipelined ADC also referred to as {\em back-end} ADC.
\begin{align}   
     V_{res} &=2V_{in}-kV_{ref}
     \label{eq:res1p5}
\end{align}
 
\subsection{Switched Capacitor Based Hadamard Transform}
\label{subsec:HT}
\begin{figure}[htbp]
\centering
\includegraphics[scale=0.125]{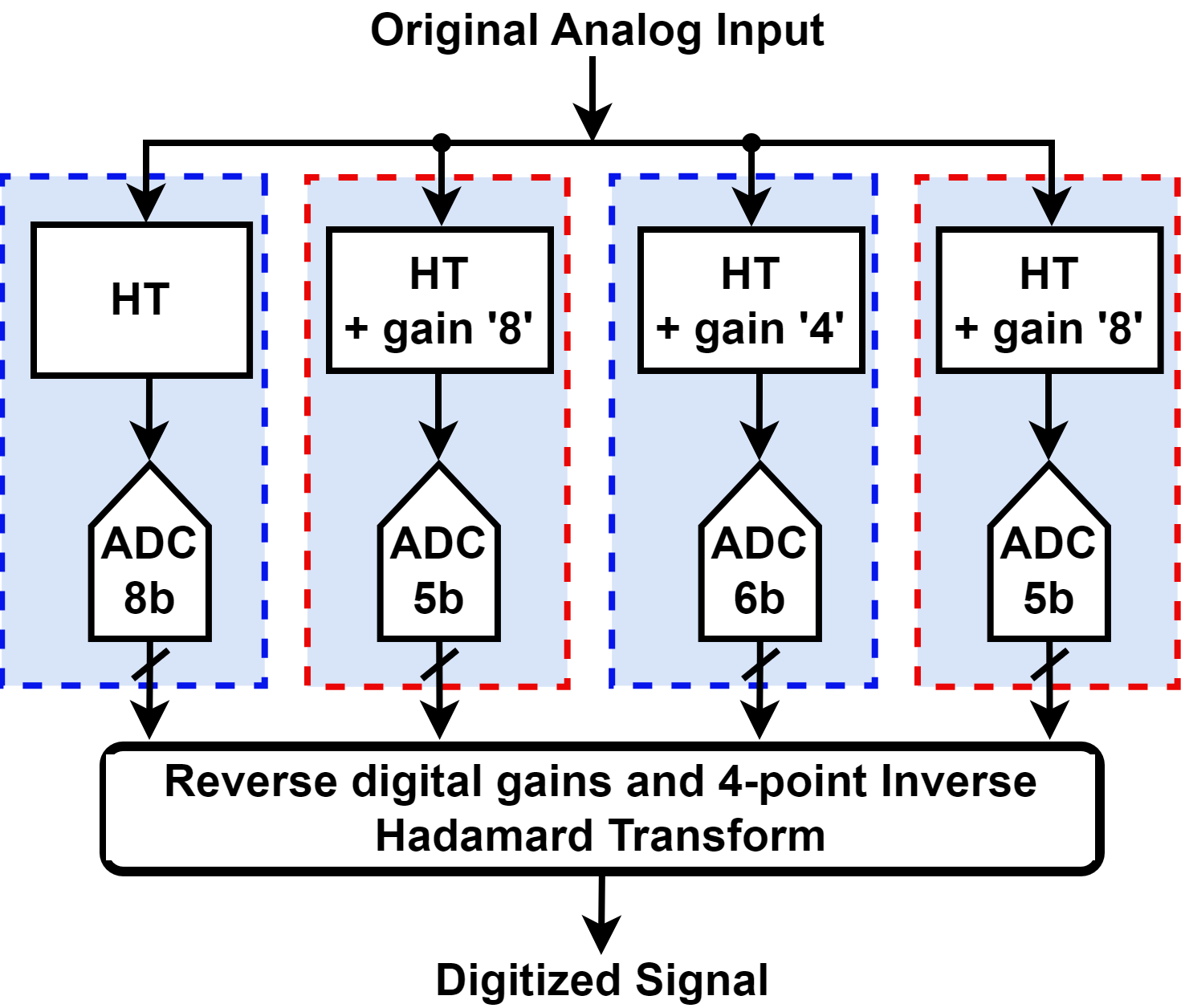}
\caption{Block Diagram illustrating a four channel ADC, where each channel includes an HT in the first stage followed by a {\em back-end} ADC.}
\label{fig: HT_BD}
\end{figure}

For an input of size `$n$' given by a vector, $V_{in}$ = $[V_{0}$, $V_{1}$, $V_{2}$,\ldots $V_{n-1}]$, the 4-point HT can be performed by taking a set of four samples at a time by each of the channels per conversion cycle. Fig. \ref{fig: HT_BD} shows the block diagram of a four channel ADC preceded by a SC based HT circuit providing requisite gain based on the ratio of $\sigma$ as discussed in Section \ref{sec: HT_and_res}.

\begin{figure}[htbp]
\centering
\includegraphics[scale=0.125]{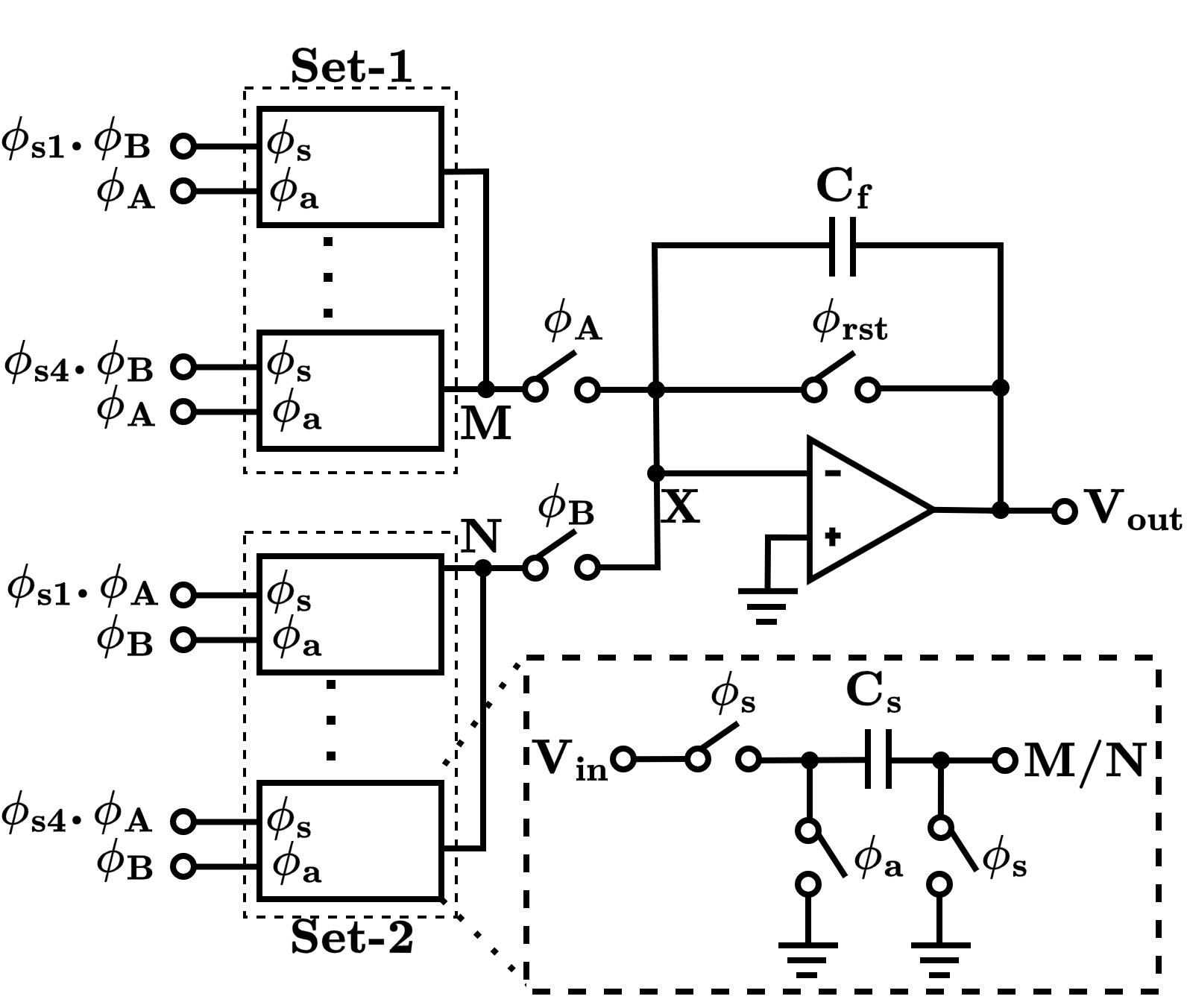}
\caption{Switch-capacitor circuit implementing HT \textcolor{black}{($\phi_a=\phi_A$ for Set-1 and $\phi_a=\phi_B$ for Set-2)}.}
\label{fig: HT-SWCAP_HT}
\end{figure}

\begin{figure}[htbp]
\centering
\includegraphics[scale=0.125]{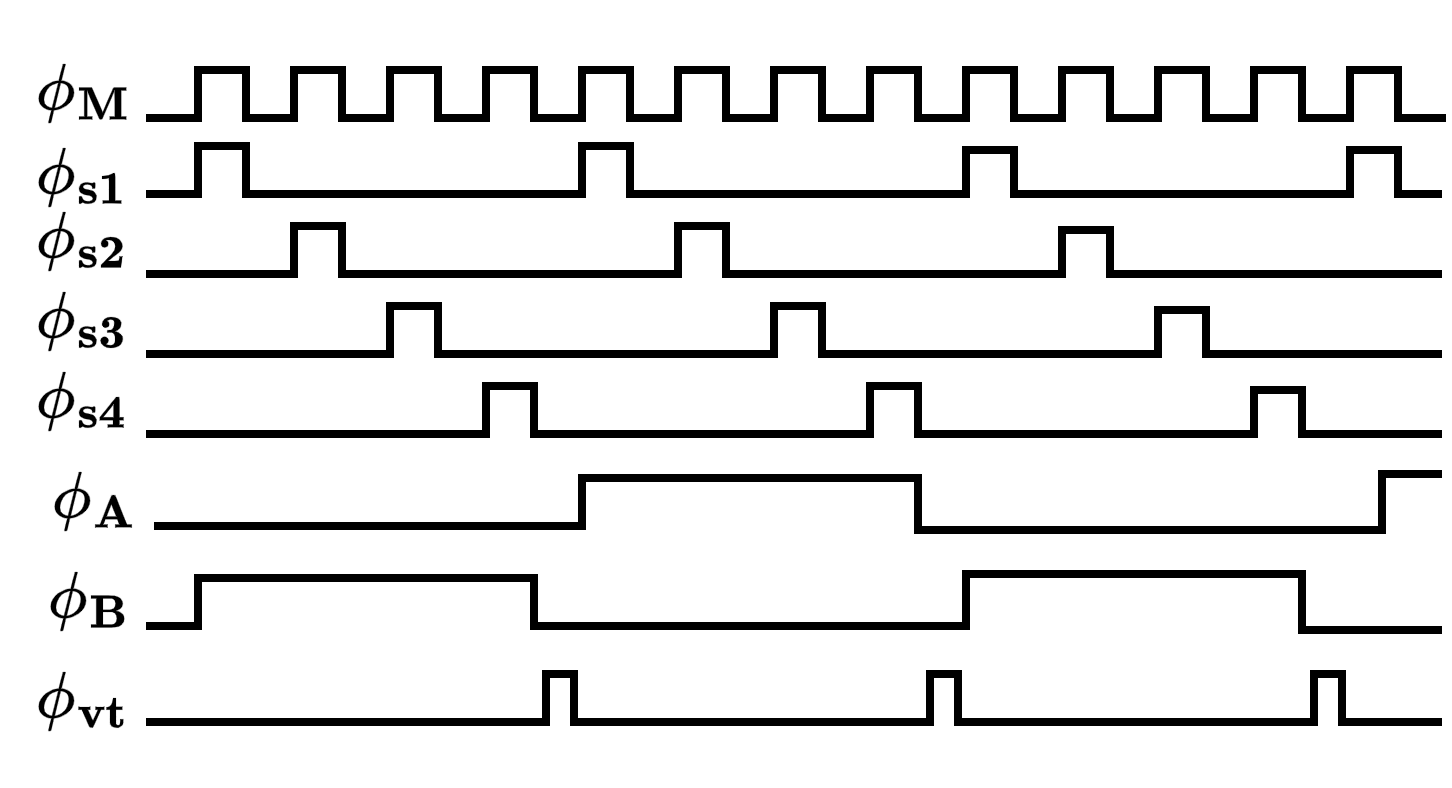}
\caption{Timing control for the HT circuits.}\vspace{-1em}
\label{fig: HT-clocks}
\end{figure}

Fig. \ref{fig: HT-SWCAP_HT} shows the SC based HT circuit. It is controlled by phase-shifted clocks derived from one master clock $\phi_{M}$ as shown in Fig. \ref{fig: HT-clocks}. The circuit consists of two sets (each with four blocks) of sampling networks controlled by a non-overlapping clock pair, $\phi_{A}$ and $\phi_{B}$. These two sets share the \textcolor{black}{op amp}, thereby operating in an interleaved manner, such that, while one set provides the output during the amplification phase, the other samples the next set of inputs. Thus, the input sampling rate is properly maintained with only one \textcolor{black}{op amp}. 

Each input image pixel is synchronized with the ON period of $\phi_{M}$. For sampling four pixels in each conversion cycle, non-overlapping shifted clocks $\phi_{s1\ldots s4}$ are used. During the $1^{st}$ ON period of clocks $(\phi_{s1\ldots s4})\cdot \phi_{B}$, first four pixels ($V_{0\ldots 3}$) are sampled onto the sampling capacitors of Set-1. As $\phi_{s4}\cdot \phi_{B}$ goes low, the total charge sampled onto the capacitors of Set-1 is $Q_{s_0}=(V_{0}+V_{1}+V_{2}+V_{3})C_{s}=V_{m_0}C_{s}$. At $\phi_{A}$, node $M$ and $X$ are connected and $Q_{s_0}$ is transferred to $C_{f}$. The output voltage is given as \textcolor{black}{$V_{out_0}=(C_sV_{m_0})/C_f$}.

During the ON period of $\phi_{A}$, the next four pixels ($V_{4\ldots 7}$) are sampled onto the sampling capacitors of Set-2 using clocks $(\phi_{s1\ldots s4})\cdot \phi_{A}$. The total sampled charge is $Q_{s_1}=(V_{4}+V_{5}+V_{6}+V_{7})C_{s}=V_{m_1}C_{s}$. When $\phi_{B}$ is high, node $N$ is connected to $X$, thus providing the voltage at the output given by \textcolor{black}{$V_{out_1}=(C_sV_{m_1})C_f$}. Thus, the non-overlapping clocks, $\phi_{A}$ and $\phi_{B}$ produce the transformed output voltage. For transforming $n$ pixels, the output of Channel-$1$ is given by (\ref{eq: HTf}), where $i=0$ to $(n/4-1)$.\vspace{-0.6em}
\begin{align}  
\label{eq: HTf}
     V_{out_i}^{ch,1} = \frac{C_{s}}{C_{f}}\sum_{i=0}^{(n/4)-1}{\left(V_{4i}+V_{4i+1}+V_{4i+2}+V_{4i+3}\right)}   
\end{align}
For channels $2$, $3$, and $4$ the outputs are given by (\ref{eq:ch2Tr}), where $\beta_j=2^{\alpha_j}/4$ for $j=2$ to $4$\textcolor{black}{, and $i=0$ to $(n/4-1)$,}  provides the requisite gain\footnote{Here the gain $\beta_j$ includes an additional scaling factor of $1/4$ so as to keep the analog values of the transformed pixels within the full-scale of the ADC.}which is based on the ratios of $\sigma$ as explained in Section \ref{sec: HT_and_res}. The subtraction operation in (\ref{eq:ch2Tr}) are realized by swapping the differential inputs\footnote{Hereafter, the expression for the three channels requiring subtraction of pixels are also shown as sum of pixels for simplicity. It is to be kept in mind that the differential implementation will make sure that subtraction is carried out.}.
\begin{align}  
      \label{eq:ch2Tr}
     \color{black}V_{out_i}^{ch,j} &\color{black}= \beta_j H_{4}^{(j)} \cdot [V_{4i},V_{4i+1},V_{4i+2},V_{4i+3}]^T
\end{align}
\textcolor{black}{where $H_4^{(j)}$ is the $j^{th}$ row of the 4-point Hadamard matrix, $H_4$.} As the HT circuit in Fig. \ref{fig: HT-SWCAP_HT} consumes power and adds thermal noise, a $1.5$-bit MDAC with embedded HT which serves as the $1^{st}$-stage of the pipelined ADC is proposed that not only saves power and area but also improves thermal noise performance as well.
\vspace{-1em}
\subsection{$1.5$-bit MDAC with Embedded HT ($1.5$-bit EHT)}
\label{subsec:1.5bHT}
\begin{figure}[htbp]
\centering
\includegraphics[scale=0.125]{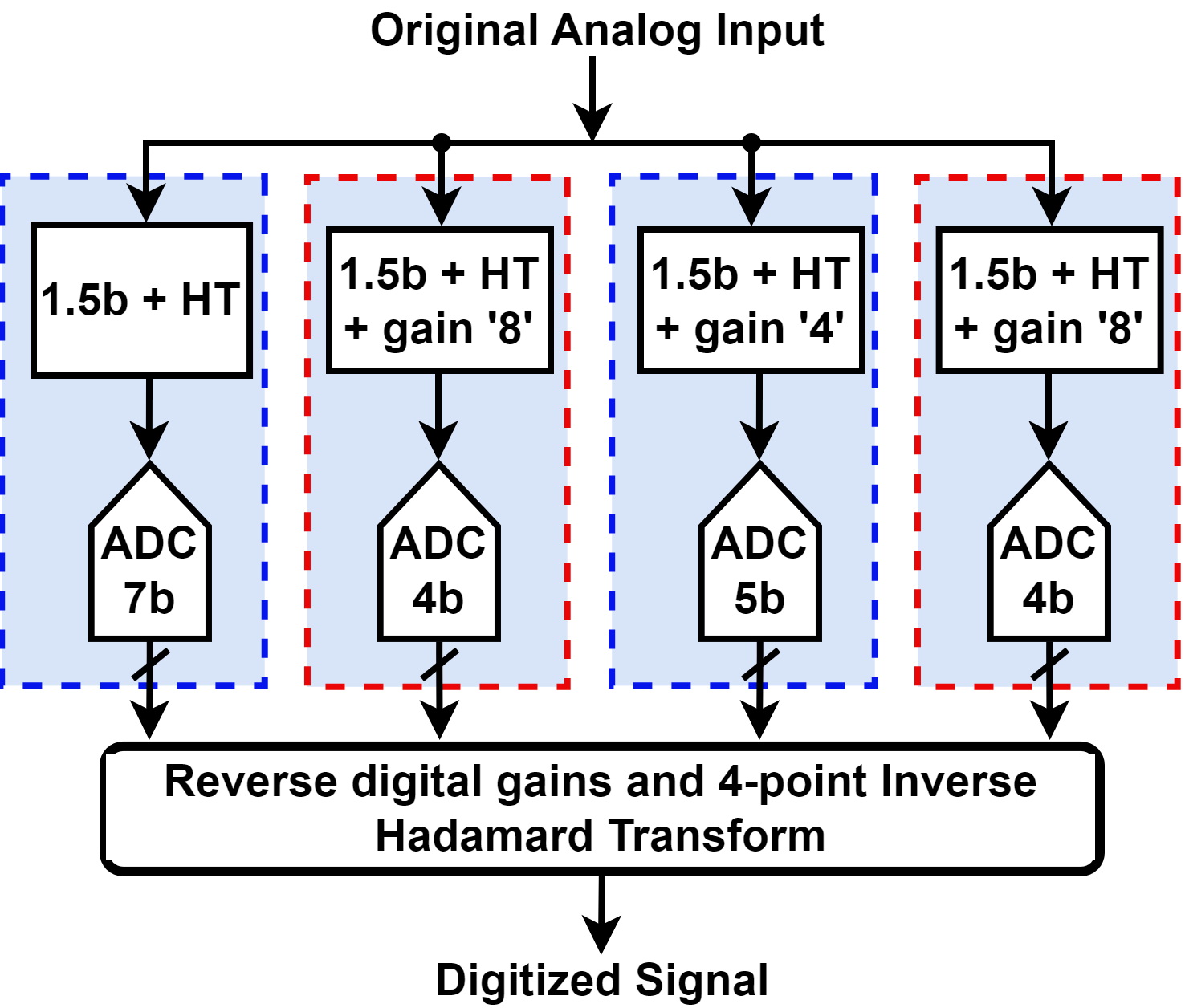}
\caption{{Block Diagram illustrating a four channel ADC, where each channel includes {\em $1.5$-bit HT gain embedded} ADC followed by a {\em back-end} ADC.}}\vspace{-1em}
\label{fig: 1.5b HT BD}
\end{figure}

Fig. \ref{fig: 1.5b HT BD} shows the proposed architecture where the $1^{st}$-stage of the pipelined ADC not only digitizes the incoming signal to $1.5$-bits but also performs the HT and hence is called the $1.5$-bit EHT stage. The output of the $1.5$-bit EHT stage is digitized by the subsequent {\em back-end} ADC which therefore requires one-bit less than the ADCs shown in Fig. \ref{fig: HT_BD}.

\begin{figure}[htbp]
\centering
\includegraphics[scale=0.125]{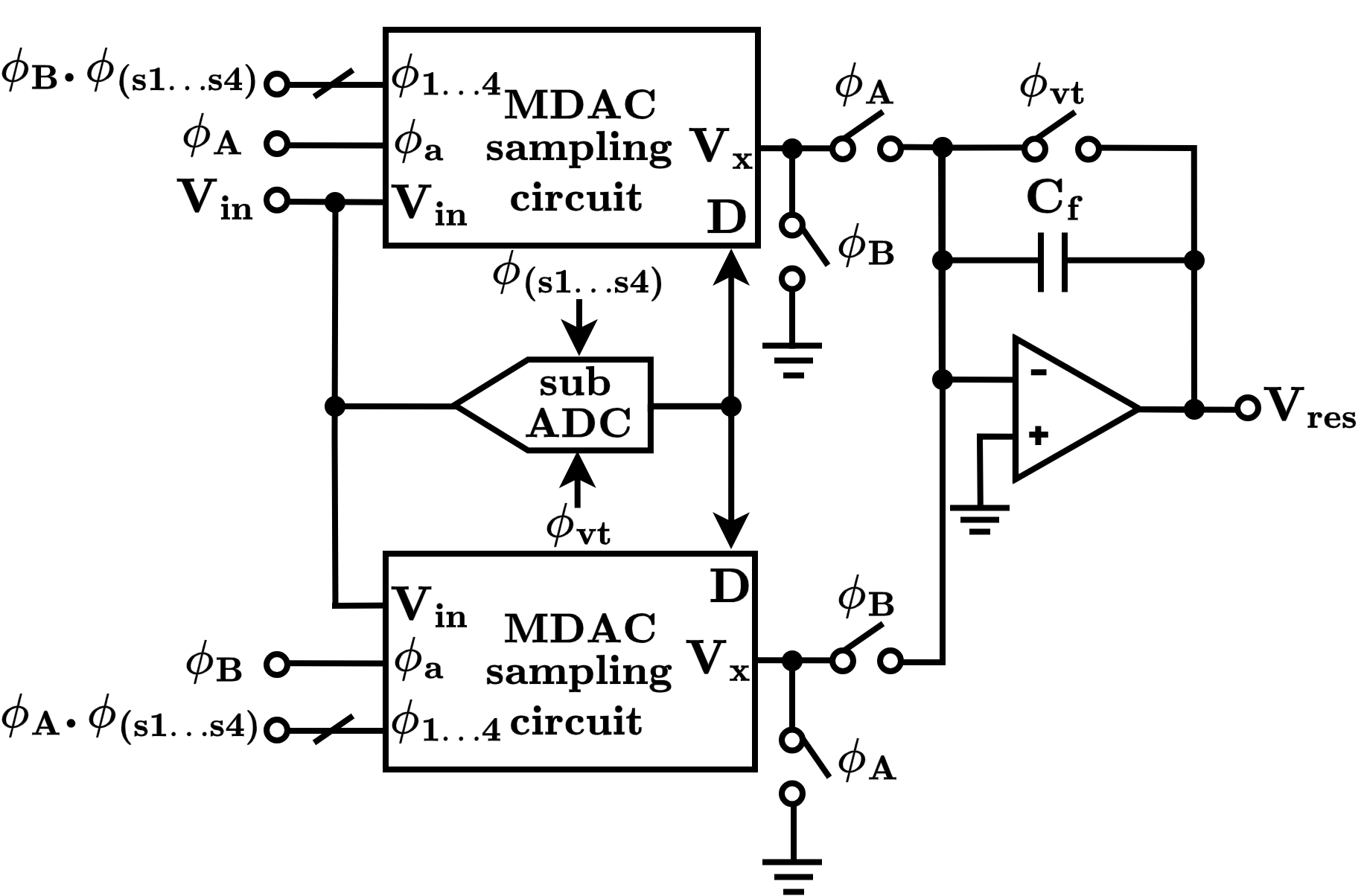}
\caption{{\em $1.5$-bit} stage with EHT.}\vspace{-1em}
\label{fig: HT-MAin}
\end{figure}

\begin{figure}[ht]
\centering
\includegraphics[scale=0.125]{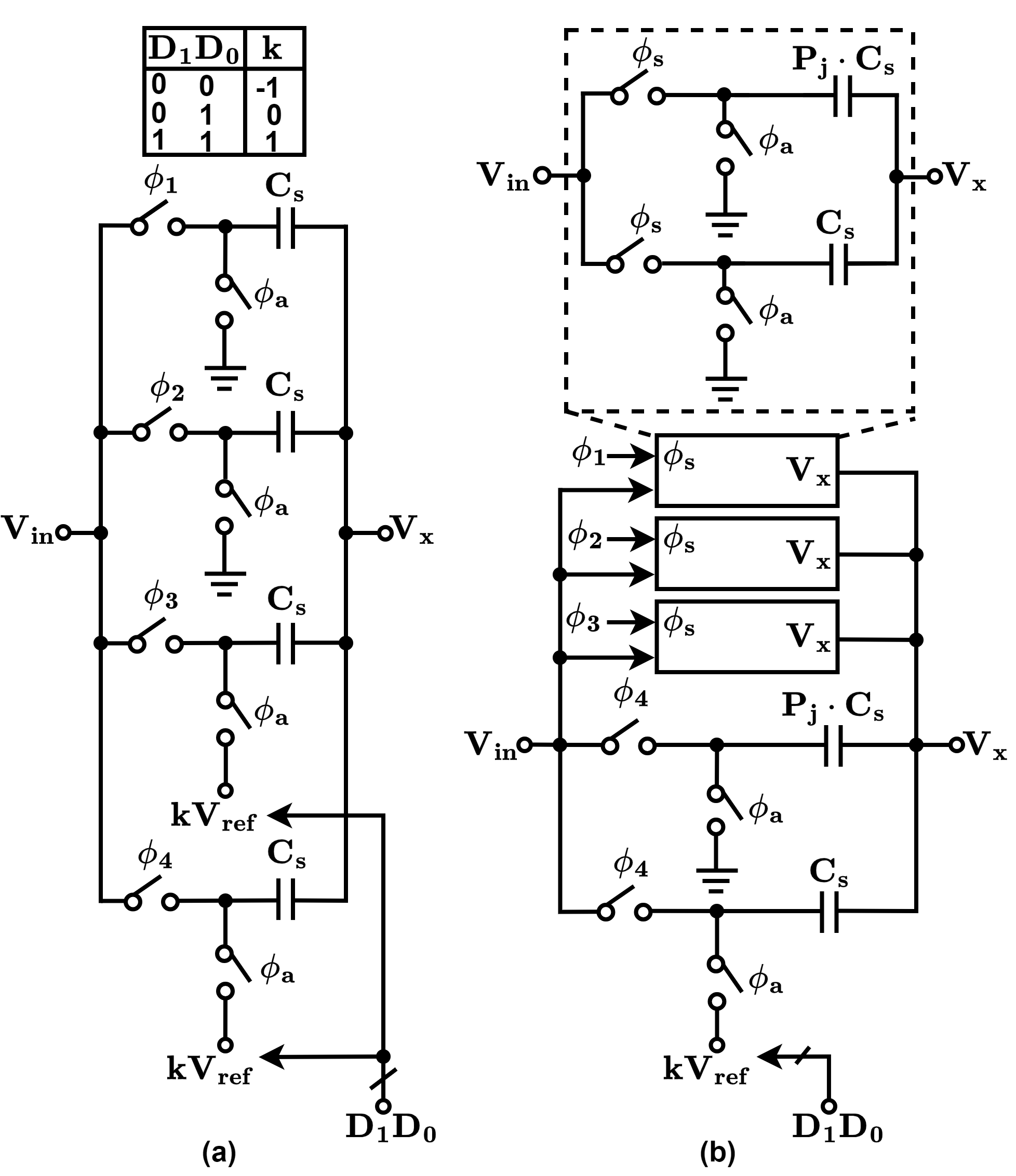}
\caption{MDAC sampling network for implementing (a) $Y_{i}$ (b) $Y_{i+1}\ldots Y_{i+3}$. Capacitors are to be scaled with $P_j=\left(2\beta_j-1\right)$ for $j=2\text{ to }4$.}
\label{fig: mdac_ch2_3and4}
\vspace{-1em}
\end{figure}

Fig. \ref{fig: HT-MAin} shows the proposed $1.5$-bit EHT stage consisting of two MDACs controlled by one sub-ADC. These blocks are obtained by modifying the $1.5$-bit MDAC and sub-ADC in a manner that perform both HT and $1.5$-bit digitization. The clocks shown in Fig. \ref{fig: HT-clocks} control the $1.5$-bit EHT. $\phi_{s1\ldots s4}$ control the sampling of the input and $\phi_{vt}$ clocks the comparators of the sub-ADC. The digital output from the sub-ADC is used to control the reference voltage to be subtracted from the input in the amplification phase of the MDAC. This residue voltage is further digitized by the {\em back-end} ADC. Thus, in Channel-$1$ for $n$ pixels the $1.5$-bit EHT should produce a residue given by (\ref{eq:HT-1}), where $V_{m_i}=\left(V_{4i}+V_{4i+1}+V_{4i+2}+V_{4i+3}\right)$. It is worth noting that $V_{m_i}$ is scaled by a factor of $4$ to ensure that the residue is bounded between $\pm V_{ref}$, thereby avoiding saturation of the {\em backend}-ADC.

\begin{align}
 V_{res_i}^{ch,1}= 2\left(\frac{V_{m_i}}{4}\right)-kV_{ref} 
 \label{eq:HT-1}
\end{align}
As mentioned in Section \ref{subsec:HT}, the other three channels of the $4$-point HT need to be amplified by `$\beta_j$'. Thus, the $1.5$-bit EHT for the remaining $3$-channels should generate a residue given by (\ref{eq:res-ch2-4}).
\begin{align}
   \color{black}V_{res_i}^{ch,j} &\color{black}= 2\beta_j H_4^{(j)} \cdot [V_{4i},V_{4i+1},V_{4i+2},V_{4i+3}]^T - kV_{ref} 
     \label{eq:res-ch2-4}
\end{align}
Fig. \ref{fig: mdac_ch2_3and4}(a) and Fig. \ref{fig: mdac_ch2_3and4}(b) show the MDAC sampling circuits to facilitate realization of (\ref{eq:HT-1}) and (\ref{eq:res-ch2-4}), respectively. Considering first four input pixels, $V_{0\ldots3}$ sampled at $(\phi_{s1\ldots s4})\cdot\phi_{B}$, the total sampled charge at the end of clock $\phi_{s4}\cdot \phi_{B}$ is $Q_{s_0}=V_{m_0}C_s$, where $V_{m_0}=(V_{0}+V_{1}+V_{2}+V_{3})$. During the amplification phase, the total charge on the capacitors is $Q_{h_0}=2kV_{ref}C_{s}+V_{res_0}C_{f}$. Since charge is conserved during the sampling and amplification phase, $Q_{s_0}= Q_{h_0}$, the residue is given by (\ref{eq:ht-res1a}),
\begin{align}
\label{eq:ht-res1a}
 \color{black}V_{res_0}^{ch,1}= \frac{C_{s}}{C_f}(V_{m_0}-kV_{ref})
\end{align}
\textcolor{black}{In (\ref{eq:ht-res1a}), $C_f=2C_s$ gives the output residue for the first channel obtained from the Hadamard transformation of the first four input pixels ($i=0$) as shown in (\ref{eq:ht-res1}). This is consistent with the expression of MDAC residue of the 1.5b EHT circuit as given by (\ref{eq:HT-1})}.
\begin{align}
V_{res_0}^{ch,1}= \frac{V_{m_0}}{2}-kV_{ref},
 \label{eq:ht-res1}
\end{align}
The two MDAC sampling networks (ref. Fig. \ref{fig: HT-MAin}), therefore, provide the residue voltage at the output of op-amp alternatively as per (\ref{eq:HT-1}).

The HT output for the rest of the three channels, {\em i.e.,} $j=2,3, \text{ and } 4$, are obtained by swapping the input lines for a differential configuration as mentioned earlier. In addition, the amplification factor $\beta_j$ is achieved by scaling the capacitors by $P_j=(2\beta_j-1)$ as shown in Fig. \ref{fig: mdac_ch2_3and4}(b). For the $1^{st}$ conversion cycle, the sampled charge at $\phi_{s4}\cdot \phi_{B}$ is given by $Q_{s_0}=V_{x_j}C_s(1+P_j)$ for $j=2$ to $4$, where $V_{x_2}=(V_{0}-V_{1}+V_{2}-V_{3})$, $V_{x_3}=(V_{0}+V_{1}-V_{2}-V_{3})$, and $V_{x_4}=(V_{0}-V_{1}-V_{2}+V_{3})$. During the amplification phase, the charge held onto the capacitors is given by $Q_{h_0}=kV_{ref}C_{s}+V_{res_0}C_{f}$. Since $Q_{s_0}=Q_{h_0}$,\vspace{-0.8em}
\begin{align}
    V_{res_0}^{ch,j}=\left[(1+P_j)V_{x_j}-kV_{ref}\right]\frac{C_s}{C_f}, \ j=2, 3, \text{ \& } 4.
    \label{eq:HT11g}
\end{align}
Putting $C_{f}=C_{s}$ and $P_j=(2\beta_j-1)$ in (\ref{eq:HT11g}), the MDAC residue expression for digitizing the first four {\em Hadamard transformed} pixels for the 3-channels is given as,
\begin{align}
 V_{res_0}^{ch,j}= 2\beta_{j}V_{x_j}-kV_{ref} \text{ for } j=2, 3, \text{ and } 4.
 \label{eq:HT-2f}
\end{align}
For transforming $n$ pixels, (\ref{eq:HT11g}) and (\ref{eq:HT-2f}) extend to (\ref{eq:HT-1}) and (\ref{eq:res-ch2-4}).

The sub-ADC comprises of two comparators whose outputs control the amplification phases of both the MDACs. For ensuring proper $1.5$-bit EHT operation, the same input voltage should get sampled onto the capacitors of both the sub-ADC and MDAC sampling network. 
\begin{figure}[ht]
\centering
\includegraphics[scale=0.125]{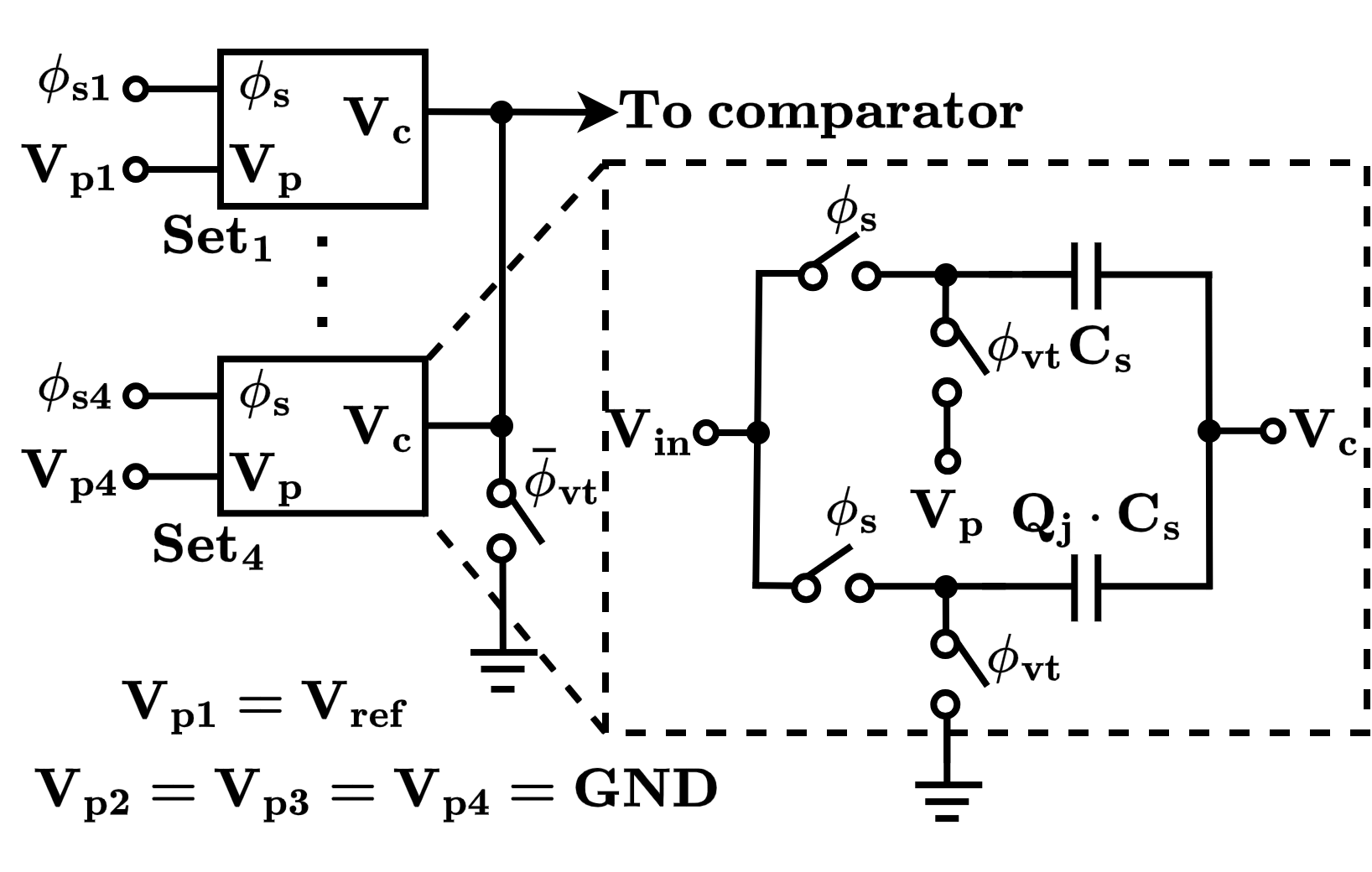}
\caption{Sub-ADC sampling network for implementing the 4-point HT channels with gain. Capacitors are to be scaled with $Q_j=\left(4\beta_j-1\right)$ for $j=1\text{ to }4$. For the first channel $Y_{i}$, $\beta_1=0.25$, thus the sampling capacitor $Q\dot C_{s}$ is not present in the circuit.}
\label{fig: sub_ch2_3and4}
\vspace{-1em}
\end{figure}

\begin{figure}[htbp]
\centering
\includegraphics[scale=0.3]{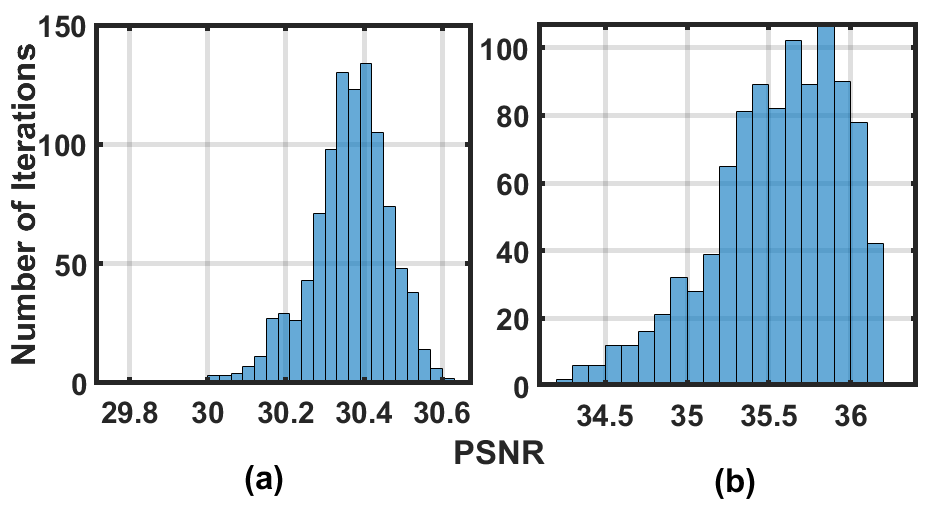} \vspace{-1em}
\caption{\textcolor{black}{Effect of 1000 iterations of 1\% random mismatch on capacitors of {\em $1.5$-bit HT embedded} circuit as in Fig. \ref{fig: 1.5b HT BD} for an image from (a) CIFAR-10 dataset (b) ImageNet dataset}.} \vspace{-1em}
\label{fig: psnr_mismatch_1.5bHT}
\end{figure}

Fig. \ref{fig: sub_ch2_3and4} shows the circuit for the sub-ADC sampling network connected to comparator's input terminal. The capacitors are scaled by a factor $Q_j=(4\beta_j-1)$ for $j=1\text{ to }4$. For channel-1, $\beta_1=0.25$, thus only four sampling capacitors are present in the circuit. The general expression for the sub-ADC of all the channels is derived as follows: During the ON period of \textcolor{black}{$\phi_{s1} \ldots \phi_{s4}$}, the input voltages $V_{0\ldots 3}$ get sampled onto the capacitors of the respective sets $Set_1 \ldots Set_4$ (ref. Fig. \ref{fig: sub_ch2_3and4}). The sampled charge on $Set_{1}$ is given by $V_{0}(1+Q_j)C_s$, $Set_{2}$ is $V_{1}(1+Q_j)C_s$, and so on. Thus, the total sampled charge onto the four sets $Set_{1\ldots 4}$ connected to the input node of comparator is $Q_{s_0}=(V_{0}+V_{1}+V_{2}+V_{3})(1+Q_j)C_s =C_{s}V_{m_0}(1+Q_j)$.\\
The decision of the comparator is made at $\phi_{vt}$, the voltage at the input of comparator is given by,
\begin{align}
   V_{m_0}C_{s}(1+Q_j)&=kV_{ref}C_{s}-4V_{c}(1+Q_j)C_{s} \nonumber \\
    \implies -4\beta_jV_{c}&=\beta_jV_{m_0}-\frac{V_{ref}}{4} \nonumber
\end{align}
Thus, the normalized input sum multiplied by the required gain is compared to $V_{ref}/4$. For $n$ pixels, the comparator node voltage is given by (\ref{eq:HTsub1}) considering total node voltage as $V^{'}_{c}=-4\beta_j\cdot V_{c}$,
\begin{align}
     V^{'ch,j}_{c_i}&=\beta_jV_{x_i}^j-\frac{V_{ref}}{4} \text{ for } j= 1, 2, 3 \text{ and } 4
    \label{eq:HTsub1}
\end{align}
Equation (\ref{eq:HTsub1}) is used by the comparators to produce a $2$-bit output, $D$, in each conversion cycle. Therefore, a $1.5$-bit EHT sub-ADC network is realized which performs the HT on the input, puts a gain and compares it to the threshold voltage to provide a digital output.\\
System level modelling of Fig. \ref{fig: 1.5b HT BD} to Fig. \ref{fig: mdac_ch2_3and4} is done by characterizing the circuits with charge equations for the SC networks. Further, non-idealities like a finite op-amp gain of $40$ dB and 1\% capacitor mismatches are introduced in the model. However, the {\em back-end} ADCs for all the channels are kept ideal. \textcolor{black}{Fig. \ref{fig: psnr_mismatch_1.5bHT}(a) shows the variation of the PSNR for 1000 iterations of a $32\times 32$ CIFAR-10 image in presence of 1\% random mismatch in the capacitors for $1000$ iterations. It can be seen that the circuit is quite robust as the PSNR only varies from 29.8dB to 30.6dB in presence of non-idealities.  Fig. \ref{fig: psnr_mismatch_1.5bHT}(b) shows a variation from 34.5dB to 36dB in case of $256\times 256$ image from ImageNet.}
\section{Measurement Setup, Results, and Discussion}
\label{sec: MM}
This section illustrates the effectiveness of the proposed idea with measurement results. Since $1.5$-bit EHT is not available as Commercial Of The Shelf (COTS) component, the overall HT based variable quantization of the image followed by inference engine is demonstrated by implementing the HT as mentioned in Section \ref{subsec:HT} on a printed circuit board (PCB) using discrete components, while being aware that this topology is {\em non-optimal}. The COTS components used for the PCB design are summarized in Table \ref{tab:pcbcomponents}.\vspace{-1em}
\begin{table}[H]
    \renewcommand{\arraystretch}{1.1}
    \centering
    \caption{Components used for the proposed HT-PCB design.}
    \label{tab:pcbcomponents}
    \begin{tabular}{|c|c|}
    \hline
    \textbf{Component} & {Part Number} \\
    \hline
    Sample and Hold & LF398 \\
    \hline
   Analog Quad CMOS Switches & DG201ACJ \\
    \hline
    Dual \textcolor{black}{Op amps} & MC33078D \\
    \hline
    D Flip Flop & SN74LS74 \\
    \hline
    Hex Schmitt Inverter & 74HC14 \\
    \hline
    Triple 3-input NAND gates & 74LS10\\
    \hline
    Potentiometers, SMD Resistors and Capacitors & \textcolor{black}{N/A} \\
    \hline
    \end{tabular} 
\end{table}\vspace{-1.5em}
\subsection{Measurement Setup and Data Acquisition}
\label{sub: PCBSche}
\begin{figure*}[htbp]
\centering
\includegraphics[scale=0.125]{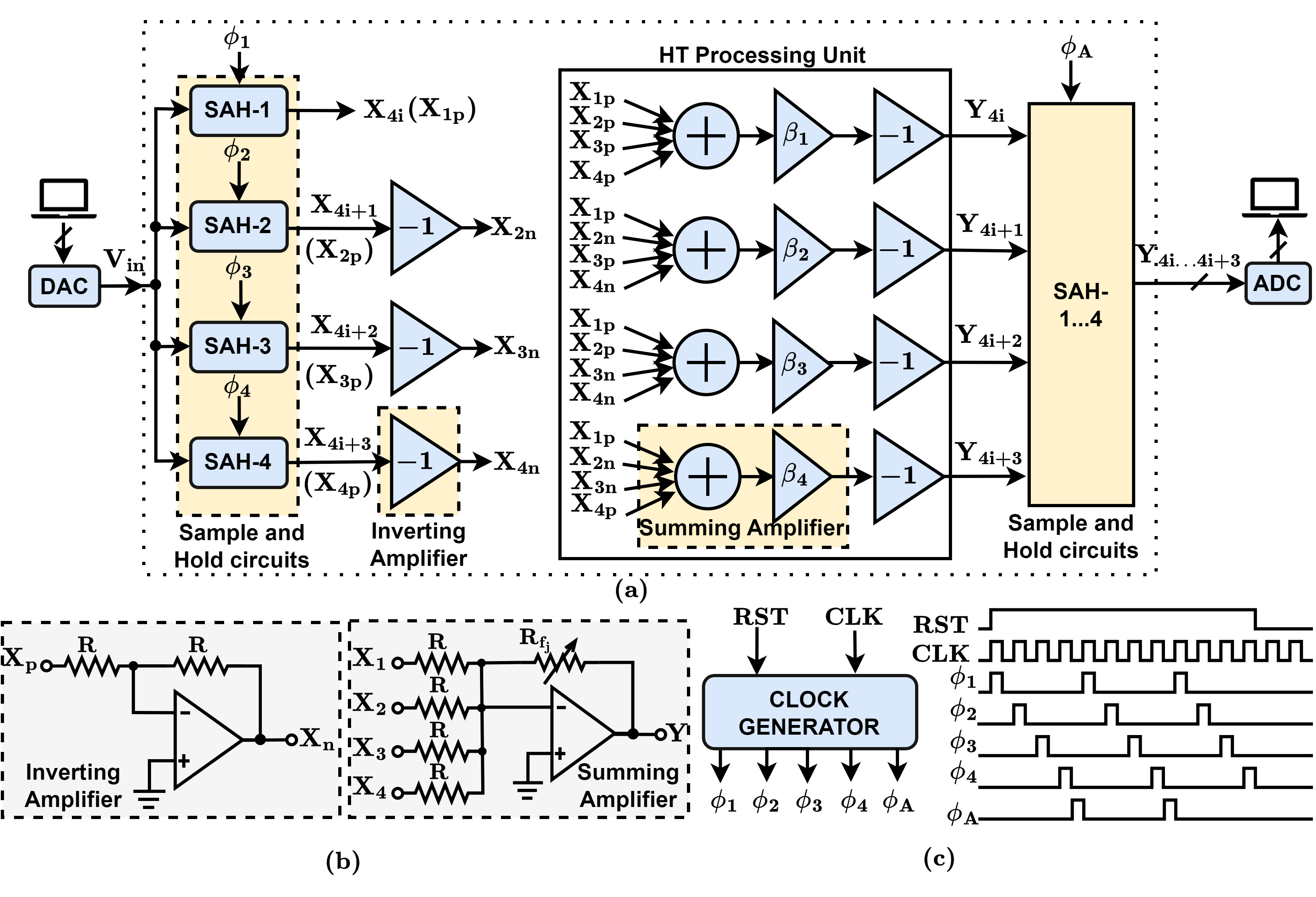}
\caption{(a) Block Diagram of the PCB implementing HT ($X_{4i\ldots 4i+3}$ are taken as $X_{1p\ldots 4p}$ for representation purpose only) (b) inverting and summing amplifier circuits to facilitate HT, and (c) clocks needed for PCB operation.}
\vspace{-1em}
\label{fig: PCB_schematic}
\end{figure*}

The PCB is tested with $100$ test images, $10$ from each of the $10$ classes of the testing set of CIFAR-10. The CIS is emulated by converting the images from CIFAR-10 dataset to analog voltage using a DAC. The block diagram in Fig. \ref{fig: PCB_schematic}(a) and Fig. \ref{fig: PCB_schematic}(b) is followed for the hardware implementation of the HT Block on PCB. The PCB takes three inputs, \emph{viz.,} the analog voltage ($V_{in}$), the main clock (CLK) and the reset signal (RST). All other clocks needed for the operation of PCB is generated by the clock generator as shown in Fig. \ref{fig: PCB_schematic}(c). The power supply needed is $\pm 5V$. The four Sample and Hold circuits (SAH) receive the analog voltage from the DAC. The SAH circuits are controlled by the multi-phase clocks $\phi_{(1\ldots 4)}$ for sampling the incoming serial data to generate four parallel channels $X_{4i}\ldots  X_{4i+3}$ (denoted as $X_{1p}\ldots X_{4p}$ in Fig. \ref{fig: PCB_schematic}(a) for simplicity of representation). At $\bar{\phi}_{A}$ \textcolor{black}{(inverted versions of $\phi_{A}$)}, analog switches pass the data to the next stage where the additive inverses of the pixels, denoted by $X_{2n\ldots 4n}$ are obtained in order to perform the subtraction operation for Channels- $2\text{ to }4$. The original pixels along with their additive inverses are passed to the HT processing unit which consists of summing amplifiers realized using \textcolor{black}{op amp}s as shown in Fig. \ref{fig: PCB_schematic}(b). The closed loop gain ($R_{f_j}/R$) of the \textcolor{black}{op amp} decides the gain $\beta_j$, where $j=1, 2, 3$ and $4$ for the respective channels. For example, in Channel-2, the required HT output is $\beta_2(X_{1p}-X_{2p}+X_{3p}-X_{4p})$. To obtain this, $X_{2n}=-X_{2p}$ and $X_{4n}=-X_{4p}$ are generated using inverting amplifiers and passed as inputs to the summing amplifier along with $X_{1p}$ and $X_{3p}$. The output from summing amplifier for channel-$2$ is given by $-\beta_2(X_{1p}+X_{2n}+X_{3p}+X_{4n})$. Since the output of the summing amplifier is inverted, an inverting amplifier with closed loop gain of $1$ is used to provide the final HT output for channel-$2$. The measurement is taken for the gains ($1, 8, 4, 8$) following the discussion in Section \ref{sec:cifar_validation}. The feedback resistors are appropriately set to obtain $\beta_j$. The four parallel output data streams are then sampled at $\phi_{A}$ by another set of SAH ICs. These parallel data are converted to a serial data stream by analog switches working at $\bar{\phi}_{(1\ldots 4)}$ \textcolor{black}{(inverted versions of $\phi_{(1\ldots 4)}$)}. The analog data stream is digitized using a 14-bit oscilloscope. Since the oscilloscope digitizes all the analog values using the same quantizer, the variable resolution is emulated by dropping the LSBs of the digitized values from the Scope using software-based quantization. For variable resolution quantization by discarding certain channels {\em viz.,} ($8,0,6,0$), ($7,0,5,0$) \emph{etc.}, the transformed pixels are completely discarded for the $2^{nd}$ and $4^{th}$ channels.

\begin{figure}[htbp]
\centering
\includegraphics[scale=0.125]{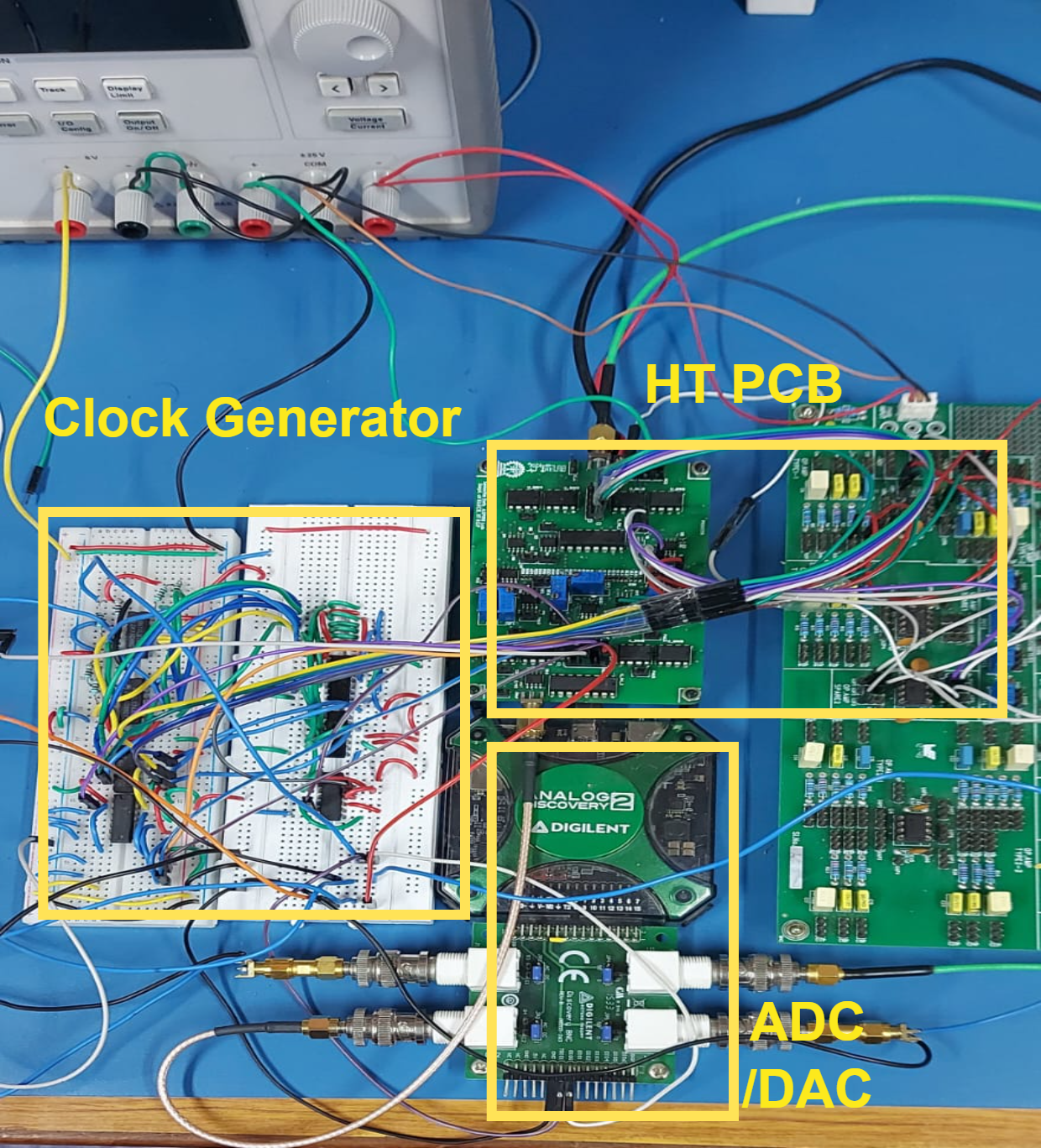}
\caption{Snapshot of the Hardware Components.} \vspace{-1em}
\label{fig: BD_PCB}
\end{figure}
\subsection{Results and Discussion}
\label{sub: MM_Discuss}
\begin{figure}[htbp]
\centering
\includegraphics[scale=0.45]{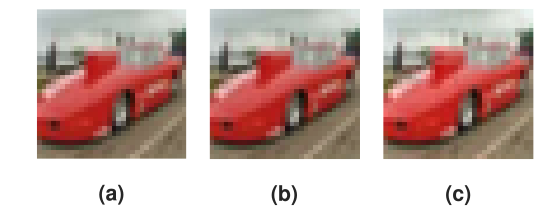}
\caption{(a) Original Image (b) Digitized Image in Simulation (c) Digitized Measured Image.} \vspace{-1em}
\label{fig: Image_comparison}
\end{figure}

The $100$ test images are first run through the software-based model to obtain the simulated results for the proposed idea. The same set of images is also passed through the PCB followed by software-based {\em variable quantization} and Inverse HT plus reverse gain to obtain the digitized images in the presence of hardware-induced errors. An example from automobile class is shown in Fig. \ref{fig: Image_comparison} with the original $8$-bit image along with proposed $6$-BPP simulated and measured images. \textcolor{black}{The PSNR and SSIM of the simulated image as compared to the original is 35dB and 0.993 respectively, whereas, the PSNR and SSIM of the measured image as compared to the original is 26dB and 0.964. Due to the little degradation in SSIM, the quality of the measured images are on par with the simulated images}.

The histogram of PSNR for the simulated and measured data for the $100$ test images for ($8,5,6,5$) and ($8,0,6,0$) are shown in Fig. \ref{fig:  psnr_100_measured}. There is some degradation in average PSNR of the measured data as compared to the simulation output for 6-BPP ($8,5,6,5$). For $3.5$-BPP ($8,0,6,0$), the measured and simulated data are close. Thus, as compared to the simulated data, the quality of measured images is not severely compromised in presence of hardware induced errors for low BPP. However, the overall image quality degrades on reducing the BPP for both simulated and measured images but this has minimal effect on the CNN performance, as is shown in this section subsequently. 

\begin{figure}[htbp]
\centering
\includegraphics[scale=0.4]{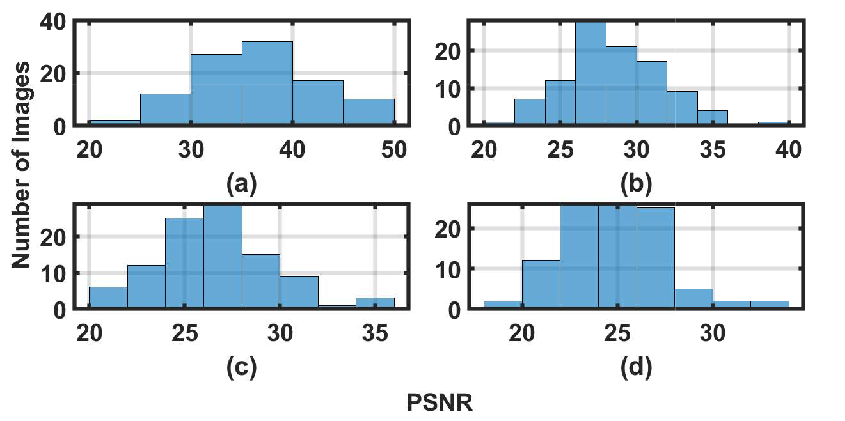}
\caption{Histogram showing PSNR for $100$ CIFAR-10 images with BPP (a) $6$-BPP ($8, 5, 6, 5$) simulated (b) $6$-BPP ($8, 5, 6, 5$) measured (c) $3.5$-BPP ($8, 0, 6, 0$) simulated (d) $3.5$-BPP ($8, 0, 6, 0$) measured.}
\label{fig: psnr_100_measured}
\end{figure}
\begin{figure}[htbp]
\centering
\includegraphics[scale=0.45]{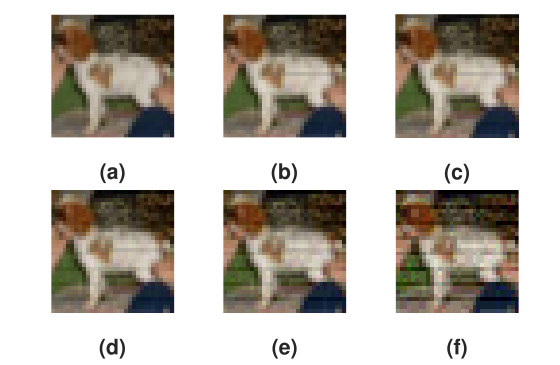}
\caption{(a) Original Image (b) 6b (c) 5b (d) 4b (e) 3b (f) 2b.}\vspace{-0.5em}
\label{fig: Image_comparison_lowres}
\end{figure}

The hardware-based results for a test image from the `Dog' class of CIFAR-10 are shown in Fig. \ref{fig: Image_comparison_lowres}. Fig.  \ref{fig: Image_comparison_lowres}(a) shows the original image input to the PCB and Fig. \ref{fig: Image_comparison_lowres}(b)-(f) show the low-resolution quantized versions of the PCB output after Inverse HT and reverse gain. The reduction in BPP is achieved by reducing the full-scale resolution. While visually there is not much difference between $5$-BPP and $4$-BPP images, the degradation starts with further lowering of the resolution. The images quantized to $2$-BPP ($4,1,2,1$) are significantly degraded and the same trend can be observed in CNN's accuracy.

Table \ref{tab:cnn_measured} shows a comparison of the results of the fine-tuned CNN mentioned in Section \ref{sub: CNN_performance} on these $100$ test images for simulated and measured data. The network gives an accuracy of $94\%$ and a loss of $1.366$ with the original $100$ test images.

\begin{table}[h]
    \renewcommand{\arraystretch}{1.1}
    \centering
    \caption{Performance comparison of CNN accuracy on 100 CIFAR-10 test images for different BPP for both  Measured and Simulated data.}
    \label{tab:cnn_measured}
    \begin{tabular}{|c|c|c|c|c|c|}
    \hline
    \multirow{2}{*}{\textbf{Bits per Channel}} & \multirow{2}{*}{\textbf{BPP}} & \multicolumn{2}{c|}{\textbf{Simulated}} & \multicolumn{2}{c|}{\textbf{Measured}} \\\cline{3-6}
    & & \textbf{Acc. (\%)} & \textbf{Loss} & \textbf{Acc. (\%)} & \textbf{Loss} \\
    \hline
    $(8, 5, 6, 5)$ & $6$ & $91$ & $1.594$ & $87$ & $2.033$ \\
    \hline
    $(7, 4, 5, 4)$ & $5$ & $90$ & $1.617$ & $87$ & $2.046$ \\
    \hline
    $(6, 3, 4, 3)$ & $4$ & $91$ & $1.699$ & $88$ & $2.075$ \\
    \hline
     $(8, 0, 6, 0)$ & $3.5$ & $91$ & $1.801$ & $90$ & $2.157$ \\
    \hline
     $(7, 0, 6, 0)$ & $3.25$ & $90$ & $1.888$ & $90$ & $2.182$ \\
    \hline
     $(7, 0, 5, 0)$ & $3$ & $89$ & $2.029$ & $90$ & $2.261$ \\
    \hline
    $(5, 2, 3, 2)$ & $3$ & $88$ & $2.145$ & $90$ & $2.326$ \\
    \hline
    $(4, 1, 2, 1)$ & $2$ & $78$ & $4.335$ & $74$ & $4.587$ \\
    \hline
    \end{tabular}
     \begin{minipage}{7.5cm}
     \vspace{0.5em}
      Note- Although the accuracy for $3$-BPP digitized measured data reaches 90\%, the loss is higher than that of the simulated data.
     \end{minipage}\vspace{-1em} 
\end{table}
The performance of network with the proposed model is similar for the measured and simulated images for resolution as low as $3$-BPP, showing that the hardware-induced errors have little effect on CNN's performance. Therefore, using the proposed method, images can be quantized to as low as $3$ bits-per-pixel while faithfully maintaining the performance of the classification algorithms.
\vspace{-0.5em}
\section{Conclusion}
\label{sec: conclusion}
This work presents a Hadamard transform-based analog transformation method for low-bit precision images. It investigates its impact on neural network performance, analyzing $5000$ ImageNet and $50000$ CIFAR-10 images. For large images, there is minimal visual degradation and no significant impact on CNN accuracy even at very low BPP. The study finds the proposed method to be better than reduced precision due to ADC bit truncation for larger images while maintaining the accuracy for lower BPP for smaller images. Using the proposed method, even with $50\%$ of transformed pixels digitized to very low BPP, negligible decrement in accuracy is observed as compared to $8$-BPP digitization. \textcolor{black}{The theoretical calculations show that there is $\approx 90\%$ power reduction, along with $\approx 50\%$ memory and I/O energy savings}. Thus, with the addition of a simple $4$ point 1D-HT to the images, low resolution digitization is feasible without impacting performance of CNN. Furthermore, a novel $1.5$-bit stage for pipelined ADC that incorporates HT along with digitization is proposed. Additionally, experimental validation with CIFAR-10 images demonstrate the effectiveness of the proposed idea, where the CNN achieves $\approx 90\%$ accuracy with only $3$-BPP. Future work could explore design and fabrication of these low power ADCs on chip level. Overall, the method achieves competitive accuracy levels across various image sizes and ADC configurations, indicating its potential for efficient hardware design for edge computing deployments without sacrificing classification accuracy.
\vspace{-0.5em}
\section*{Acknowledgements}
The authors would like to thank Anindya Sundar Dhar, from the Department of Electronics and Electrical Communication Engineering, IIT Kharagpur for his invaluable inputs and discussion towards the development of PCB and the test setup.
\bibliographystyle{IEEEtran} \vspace{-1em}
\bibliography{references}
\end{document}